\documentclass[11pt]{article}
\usepackage{}
\usepackage{amssymb}
\usepackage{amsfonts}
\usepackage{mathrsfs}
\usepackage{graphicx}
\usepackage{amsmath}
\usepackage{color}
\usepackage{amsthm}
\usepackage{epstopdf}
\usepackage{subfigure}
\usepackage{pifont,bm}
\usepackage{tikz}
\usepackage{enumerate}
\usepackage{mathrsfs}
\usepackage{cite}
\usepackage{hyperref}

\theoremstyle{definition}

\makeatletter
\def\@biblabel#1{[#1]}
\makeatother

\makeatletter \@addtoreset{equation}{section}

\begin{document}

\begin{titlepage}
\title{\bf{The data-driven localized wave solutions of the derivative nonlinear
Schr\"{o}dinger equation by using improved PINN approach
\footnote{Corresponding authors.\protect\\
\hspace*{3ex} E-mail addresses: ychen@sei.ecnu.edu.cn (Y. Chen)}
}}
\author{Juncai Pu$^{a}$, Weiqi Peng$^{a}$, Yong Chen$^{a,b,*}$\\
\small \emph{$^{a}$School of Mathematical Sciences, Shanghai Key Laboratory of Pure Mathematics and} \\
\small \emph{Mathematical Practice, East China Normal University, Shanghai, 200241, China} \\
\small \emph{$^{b}$College of Mathematics and Systems Science, Shandong University }\\
\small \emph{of Science and Technology, Qingdao, 266590, China} \\
\date{}}
\thispagestyle{empty}
\end{titlepage}
\maketitle

\vspace{-0.5cm}
\begin{center}
\rule{15cm}{1pt}\vspace{0.3cm}

\parbox{15cm}{\small
{\bf Abstract}\\
\hspace{0.5cm}
The research of the derivative nonlinear Schr\"odinger equation (DNLS) has attracted more and more extensive attention in theoretical analysis and physical application. The improved physics-informed neural network (IPINN) approach with neuron-wise locally adaptive activation function is presented to derive the data-driven localized wave solutions, which contain rational solution, soliton solution, rogue wave, periodic wave and rogue periodic wave for the DNLS with initial and boundary conditions in complex space. Especially, the data-driven periodic wave and  rogue periodic wave of the DNLS are investigated by employing the IPINN method for the first time. Furthermore, the relevant dynamical behaviors, error analysis and vivid plots have been exhibited in detail. The numerical results indicate the IPINN method can well simulate the localized wave solutions of the DNLS under a small data set.

}

\vspace{0.5cm}
\parbox{15cm}{\small{

\vspace{0.3cm} \emph{Key words:The data-driven localized wave solutions; The derivative nonlinear Schr\"odinger equation; Improved physics-informed neural networks}  \\

} }
\end{center}
\vspace{0.3cm} \rule{15cm}{1pt} \vspace{0.2cm}

\section{Introduction}
With the revolution of hardware technology, the great improvement of computer speed and the explosive growth of available data, promoting the application of machine learning and data analysis in practice has made very important progress in pattern recognition, natural language processing, computer vision, cognitive science, genomics and many other fields \cite{LeCun2015,Bishop2006,Collobert2011,Krizhevsky2017,Lake2015,Alipanahi2015}. As is known to all, neural networks (NNs) is an extensive parallel interconnected network composed of adaptive simple units, its organization can simulate the interaction of biological neural system to real world objects \cite{Kohonen1988}. Furthermore, the deep learning of multilayer NNs can solve many practical problems, it has attracted more and more attention in recent years \cite{Krizhevsky2017,Lake2015,Alipanahi2015}. However, in the process of analyzing complicated mathematical, physical, biological and engineering systems, the cost of data acquisition is generally too high, how to utilize machine learning approach to draw conclusions and make decisions under the small data regime is a significant practical problem. In addition, under the environment with only partial information, the vast majority of state-of-the-art machine learning technologies, such as convolution and recurrent NNs \cite{LeCun1989,Rumelhart1986}, lack robustness and fail to provide any guarantee of convergence. Recently, a new NNs which also be called physics-informed neural networks(PINNs) has been proposed and proved to be particularly suitable for solving and inversing equations which have been controlled via mathematical physical systems based on the deep learning models of multilayer NNs, and found that the high-dimensional network tasks can be completed with less data sets\cite{Raissi2019}. That is the PINNs method with the small data regime can not only accurately solve both forward problems, where the approximate solutions of governing equations are obtained, but also precisely deal with the highly ill-posed inverse problems, where parameters involved in the governing equation are inferred from the training data. Subsequently, in order to improve the convergence rate and training effect for PINNs, Jagtap and collaborators presented two different kinds of adaptive activation functions, namely global adaptive activation functions and locally adaptive activation functions, to approximate smooth and discontinuous functions as well as solutions of linear and nonlinear partial differential equations, and introduced a scalable parameters in the activation function, which can be optimized to achieve best performance of the network as it changes dynamically the topology of the loss function involved in the optimization process \cite{Jagtap2020,JagtapA2020}. Moreover, compared with global adaptive activation functions, the numerical results demonstrate the locally adaptive activation functions further improve the training speed and performance of NNs \cite{JagtapA2020}. Furthermore, a slope recovery term based on activation slope has been added to the loss function of locally adaptive activation functions to further improve the performance and speed up the training process of NNs.

The derivative nonlinear Schr\"{o}dinger equation (DNLS) plays a significant role both in the integrable system theory and many physical applications, especially in space plasma physics and nonlinear optics \cite{Kaup1978,Mjolhus1976}. In 1976, Mio and co-workers first derived the DNLS from Alfv\'{e}n wave propagation in plasma, and found it well described the propagation of small amplitude nonlinear Alfv\'{e}n wave in low plasma \cite{Mio1976}. Furthermore, it was shown that the DNLS can describe the behaviour of large-amplitude magnetohydrodynamic waves propagating in an arbitrary direction with respect to the magnetic field in a high-$\beta$ plasma as well \cite{Ruderman2002}. In nonlinear optics, the DNLS also describes the transmission of sub-picosecond pulses in single mode optical fibers \cite{Tzoar1981}, and the DNLS can be derived in the theory of ultrashort femtosecond nonlinear pulse in optical fiber \cite{ChenX2004}. In electromagnetism, the filamentation of lower-hybrid waves can be simulated by the DNLS which governs the asymptotic state of the filamentation, and it admits moving solitary envelope solutions for the electric field \cite{Spatchek1977}. Therefore, it is very significant to find abundant solutions of the DNLS for explaining various complex physical phenomena and revealing more unknown physical laws.

For decades, some classical solutions and important results of the DNLS have been obtained with the aid of different approaches. In 1978, Kaup and Newell demonstrated the integrability of the DNLS in the sense of inverse scattering method \cite{Kaup1978}. Furthermore, with the aid of the Hirota bilinear method, the $N$-soliton formula of the DNLS has been first constructed by Nakamura and Chen \cite{Nakamura1980}. According to the Darboux transform approach, Huang and Chen derived the determinant form of $N$-soliton formula \cite{Huang1990}. Kamchatnov and cooperators not only proposed a method for constructing periodic solutions of several integrable evolution equations and applied it to the DNLS, but also found the formation of solitons on the sharp front of optical pulse in an optical fiber via the DNLS \cite{Kamchatnov1990,Kamchatnov1998}. Moreover, Hayashi and Ozawa discussed the Cauchy problem of the DNLS in detail \cite{Hayashi1992}. The compact $N$-soliton formulae both with asymptotically vanishing and non-vanishing amplitudes were pursued by iterating B\"acklund transformation of the DNLS \cite{Steudel2003}. Recently, various methods have been utilized to reveal more abundant solutions and more new physical phenomena of the DNLS \cite{Guo2012,XuT2018,Xue2020,Xu2019,Zhang2020,Xu2011,Yang2020,ZhangY2014,Liu2018,ChenJB2021}. Although correlative methods and theories of explicit solutions of DNLS have become increasingly mature, it is still a difficult problem to obtain the solutions of DNLS via numerical method with small data samples.

Recently, recovering the data-drive solutions and revealing the dynamic behaviors of nonlinear partial differential equations with physical constraints have attracted extensive attention and set off a research boom by using the PINN method \cite{Raissi2019}. Due to the abundant sample space and good properties of integrable systems, it has become a research hotspot to apply the PINN approach to the field of integrable systems, and an effective numerical calculation method and complete theory will be established. More recently, Chen research team constructed a great quantity data-driven solutions for many nonlinear integrable systems by using PINN deep learning method \cite{Li2020,LiJ2020,Pu2021,PuJ2021,Peng2021}. Moreover, the dynamic behaviors and data-driven solutions of some other nonlinear systems have also been studied by using PINN method \cite{Wang2021}. Especially, the solitons, breathers and rogue wave solutions of the nonlinear Schr\"odinger equation have been recovered with the help of the PINN model \cite{Pu2021}. Furthermore, an improved PINN (IPINN) approach with neuron-wise locally adaptive activation function was presented to derive rational soliton solutions and rogue wave solutions of the DNLS in complex space, and numerical results demonstrated the improved approach has faster convergence and better simulation effect than classical PINN method \cite{PuJ2021}. In this paper, we will consider localized wave solutions which contain periodic wave solution, rogue wave solution,  rogue periodic wave solution and rational solutions of the DNLS by utilizing the IPINN approach with neuron-wise locally adaptive activation function \cite{JagtapA2020,PuJ2021}. We focus on the following DNLS with initial-boundary value conditions, whose dispersion term is different from that of the DNLS in Ref. \cite{PuJ2021}, the expression is as follows
\begin{align}\label{E1}
\begin{split}
\begin{cases}
q_t+\mathrm{i}q_{xx}+(|q|^2q)_x=0,\,x\in[L_0,L_1],\, t\in[T_0,T_1],\\
q(x,T_0)=q_0(x),\,x\in[L_0,L_1],\\
q(L_0,t)=q_{\mathrm{lb}}(t),\,q(L_1,t)=q_{\mathrm{ub}}(t),\,t\in[T_0,T_1],\\
\end{cases}
\end{split}
\end{align}
where the subscripts denote the partial derivatives of the complex field $q(x,t)$ with respect to the space $x$ and time $t$, and the $L_0$ and $L_1$ represent the lower and upper boundaries of $x$ respectively. Similarly, $T_0$ and $T_1$ represent the initial and final times of $t$ respectively. Moreover, the $q_{\mathrm{lb}}(t)$ and $q_{\mathrm{ub}}(t)$ are the lower and upper boundaries of the $q(x,t)$ corresponding to $x=L_0$ and $x=L_1$ respectively.

This paper is organized as follows. In section 2, we introduce briefly discussions of the IPINN method with locally adaptive activation function, where also discuss about training data, loss function, optimization method and the operating environment. Moreover, the algorithm flow schematic of the DNLS based on IPINN model is exhibited in detail. In Section 3, the data-driven rational soliton, rational phase solution and vivid plots of DNLS have been exhibited via IPINN model. Section 4 provides the periodic wave, rogue wave and rogue periodic wave of the DNLS by utilizing the improved PINN approach, and related plots and dynamic analysis are given out in detail. Conclusion is given out in last section.

\section{The improved PINN method}

In general, we consider the general (1+1)-dimensional nonlinear time-dependent equations in complex space, in which each contains a dissipative term as well as other partial derivatives, such as nonlinear terms or dispersive terms, as shown below
\begin{align}\label{E2}
q_t+\mathcal{N}(q,q_x,q_{xx},q_{xxx},\cdots)=0,
\end{align}
where $q$ are complex-valued solutions of $x$ and $t$ to be determined later, and $\mathcal{N}$ is a nonlinear functional which contains the solution $q(x,t)$, its derivatives of arbitrary order respecting to $x$ and any combination between them. Due to the complexity of the structure of the complex-valued solutions $q(x,t)$ in Eq. \eqref{E2}, we decompose $q(x,t)$ into the real part $u(x,t)$ and the imaginary part $v(x,t)$ by employing two real-valued functions $u(x,t)$ and $v(x,t)$, that is $q(x,t)=u(x,t)+\mathrm{i}v(x,t)$. Then substituting it into Eq. \eqref{E2}, we have
\begin{align}\label{E3}
u_t+\mathcal{N}_u(u,u_x,u_{xx},u_{xxx},\cdots)=0,
\end{align}
\begin{align}\label{E4}
v_t+\mathcal{N}_v(v,v_x,v_{xx},v_{xxx},\cdots)=0.
\end{align}
Similarly, the $\mathcal{N}_u$ and $\mathcal{N}_v$ are nonlinear functionals which consist of the corresponding solutions, their derivatives of arbitrary order respecting to $x$ and any combination between them, respectively. Then the physics-informed neural networks $f_u(x,t)$ and $f_v(x,t)$ can be defined as
\begin{align}\label{E5}
f_u:=u_t+\mathcal{N}_u(u,u_x,u_{xx},u_{xxx},\cdots),
\end{align}
\begin{align}\label{E6}
f_v:=v_t+\mathcal{N}_v(v,v_x,v_{xx},v_{xxx},\cdots).
\end{align}

The original PINN method could not accurately reconstruct solutions of complex forms in some complicated nonlinear equations. Therefore, due to some accuracy and performance requirements, we draw into an IPINN where a neuron-wise locally adaptive activation function technique is introduced into the classical PINN method in this paper. It changes the slope of the activation function adaptively, resulting in non-vanishing gradients and faster training of the network. Specifically, we first define such neuron-wise locally adaptive activation function as
\begin{align}\nonumber
\sigma\left(na^d_i\left(\mathcal{L}_d\left(\textbf{x}^{d-1}\right)\right)_i\right),d=1,2,\cdots,D-1,i=1,2,\cdots,N_d,
\end{align}
where $n>1$ is a scaling factor and $\{a^d_i\}$ are additional $\sum\limits_{d=1}^{D-1}N_d$ parameters to be optimized. Note that, there is a critical scaling factor $n_{c}$, and the optimization algorithm will become sensitive when $n\geqslant n_c$ in each problem set. The neuron activation function acts as a vector activation function in each hidden layer, and each neuron has its own slope of activation function.

The new NN with neuron-wise locally adaptive activation function can be represented as
\begin{align}\label{E7}
q(\textbf{x};\bar{\Theta})=\left(\mathcal{L}_D\circ\sigma\circ na^{D-1}_{i}\left(\mathcal{L}_{D-1}\right)_{i}\circ\cdots\circ\sigma\circ na^1_i\left(\mathcal{L}_1\right)_i\right)(\textbf{x}),
\end{align}
where the set of trainable parameters $\bar{\Theta}\in\bar{\mathcal{P}}$ consists of $\big\{\textbf{W}^d,\textbf{b}^d\big\}_{d=1}^{D}$ and $\big\{a_i^d\big\}_{i=1}^{D-1},\forall i=1,2,\cdots,N_d$, $\bar{\mathcal{P}}$ is the parameter space. In this method, the initialization of scalable parameters are carried out in the case of $na_i^d=1,\forall n\geqslant1$.

The resulting optimization algorithm will attempt to find the optimized parameters including the weights, biases and additional coefficients in the activation to minimize the new loss function defined as
\begin{align}\label{E8}
Loss=Loss_q+Loss_f+Loss_a,
\end{align}
where $Loss_q, Loss_f$ are defined as following
\begin{align}\label{E9}
Loss_q=\frac{1}{N_q}\left[\sum^{N_q}_{i=1}|u(x_u^i,t_u^i)-u^i|^2+\sum^{N_q}_{i=1}|v(x_v^i,t_v^i)-v^i|^2\right],
\end{align}
and
\begin{align}\label{E10}
Loss_f=\frac{1}{N_f}\left[\sum^{N_f}_{j=1}|f_u(x_f^j,t_f^j)|^2+\sum^{N_f}_{j=1}|f_v(x_f^j,t_f^j)|^2\right],
\end{align}
where $\{x^i_u,t^i_u,u^i\}^{N_q}_{i=1}$ and $\{x^i_v,t^i_v,v^i\}^{N_q}_{i=1}$ denote the initial and boundary value data on Eqs. \eqref{E3} and \eqref{E4}. Furthermore, $\{x_f^j,t_f^j\}^{N_{f}}_{j=1}$ represent the collocation points on networks $f_u(x,t)$ and $f_v(x,t)$. The last slope recovery term $Loss_a$ in the loss function \eqref{E6} is defined as
\begin{align}\label{E11}
Loss_a=\frac{1}{\frac{N_a}{D-1}\sum\limits_{d=1}^{D-1}\mathrm{exp}\Bigg(\frac{\sum\limits_{i=1}^{N_d}a_i^d}{N_d}\Bigg)},
\end{align}
where $N_a$ has been imposed to control the range of the value size of $Loss_a$, and we all take $N_a=10$ for dominating the loss function and ensuring that the loss value is not too large in this paper. Here, term $Loss_a$ forces the NN to increase the activation slope value quickly, which ensures the non-vanishing of the gradient of the loss function and improves the network's training speed. Consequently, $Loss_q$ corresponds to the loss on the initial and boundary data, the $Loss_f$ penalizes the DNLS not being satisfied on the collocation points, and the $Loss_a$ changes the topology of $Loss$ function and improves the convergence speed and network optimization ability.

\begin{figure}[htbp]
\centering
\begin{minipage}[t]{0.99\textwidth}
\centering
\includegraphics[height=9cm,width=15cm]{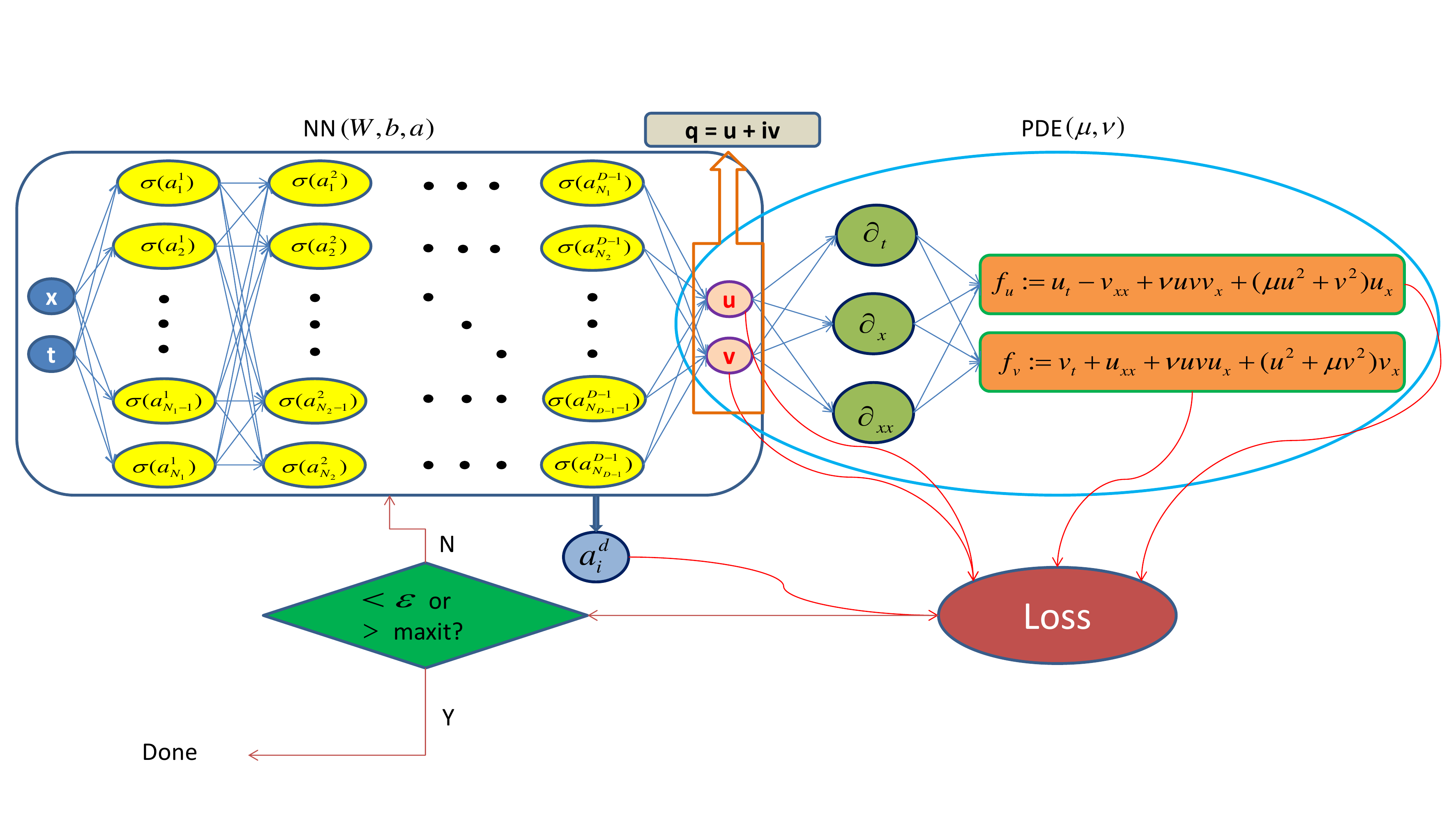}
\end{minipage}
\centering
\caption{(Color online) Schematic of IPINN for the DNLS. The left NN is the uninformed network while the right one induced by the governing equation is the informed network. The two NNs share hyper-parameters and they both contribute to the loss function.}
\label{F1}
\end{figure}

In order to understand IPINN approach more clearly, the IPINN algorithm flow model of the DNLS is shown in following. Fig.\ref{F1} gives a sketch of IPINN algorithm for the DNLS where one can see the NN along with the supplementary physics-informed part. The loss function is evalu-ated using the contribution from the NN part as well as the residual from the governing equation given by the physics-informed part. Then, one seeks the optimal values of weights $\textbf{W}$, biases $\textbf{b}$ and scalable parameter $a^d_i$ in order to minimize the loss function below certain tolerance $\varepsilon$ until a prescribed maximum number of iterations.

In this method, all loss functions are simply optimized by employing the L-BFGS algorithm, which is a full-batch gradient descent optimization algorithm based on a quasi-Newton method \cite{Liu1989}. Especially, the scalable parameters in the adaptive activation function are initialized generally as $n=5,a_i^d=0.2$, unless otherwise specified. In addition, we select relatively simple multi-layer perceptrons (i.e., feedforward NNs) with the Xavier initialization and the hyperbolic tangent ($\tanh$) as activation function. All the codes in this article is based on Python 3.7 and Tensorflow 1.15, and all numerical experiments reported here are run on a DELL Precision 7920 Tower computer with 2.10 GHz 8-core Xeon Silver 4110 processor and 64 GB memory.

\section{The data-driven rational solution and soliton solution of the DNLS}
In this section, we numerically reveal two different types of solutions and their corresponding dynamic analysis for the DNLS by using the IPINN which contains nine hidden layers with each layer having 40 neurons. The accurate rational solution and soliton solution have been obtained by using the Darboux transformation \cite{Xu2019}.

\subsection{The data-driven rational solution}
It is known that Eq. \eqref{E1} admits the explicit rational solution \cite{Xu2019}
\begin{align}\label{E12}
q_{\mathrm{rs}}(x,t)=\frac{4\mathrm{e}^{2\mathrm{i}(2t-x)}(4\mathrm{i}(4t-x)-1)^3}{(16(4t-x)^2+1)^2}.
\end{align}

In what follows, we will consider the initial condition $q_{\mathrm{rs}}(x,T_0)$ and Dirichlet boundary condition $q_{\mathrm{rs}}(L_0,t)$ and $q_{\mathrm{rs}}(L_1,t)$ of Eq. \eqref{E1} arising from the rational solution Eq. \eqref{E12}. Here we take $[L_0,L_1]$ and $[T_0,T_1]$ in Eq. \eqref{E1} as $[-5.0,5.0]$ and $[-0.08,0.08]$, respectively. Then, we focus on the corresponding the Cauchy problem with initial condition $q_0(x)$, as shown below
\begin{align}\label{E13}
q_{\mathrm{rs}}(x,-0.08)=\frac{4\mathrm{e}^{2\mathrm{i}(-0.16-x)}(4\mathrm{i}(-0.32-x)-1)^3}{(16(-0.32-x)^2+1)^2},\,x\in[-5.0,5.0],
\end{align}
and the Dirichlet boundary conditions
\begin{align}\label{E14}
q_{\mathrm{lb}}(t)=q_{\mathrm{rs}}(-5.0,t),\,q_{\mathrm{ub}}(t)=q_{\mathrm{rs}}(5.0,t),\,t\in[-0.08,0.08].
\end{align}

We employ the traditional finite difference scheme on even grids in MATLAB to simulate Eq. \eqref{E12} which contains the initial data \eqref{E13} and boundary data \eqref{E14} to acquire the training data. Specifically, divide spatial region $[-5.0,5.0]$ into 513 points and temporal region $[-0.08,0.08]$ into 401 points, rational solution \eqref{E12} is discretized into $401$ snapshots accordingly. We generate a smaller training dataset containing initial-boundary data by randomly extracting $N_q=400$ from original dataset and $N_f=20000$ collocation points which are generated by the Latin hypercube sampling method (LHS) \cite{Stein1987}. After giving a dataset of initial and boundary points, the latent rational solution $q(x,t)$ has been successfully learned by tuning all learnable parameters of the IPINN and regulating the loss function \eqref{E8}. The model of IPINN achieves a relative $\mathbb{L}_2$ error of 6.042053$\mathrm{e}-$02 in about 1159.0106 seconds, and the number of iterations is 7448.

\begin{figure}[htbp]
\centering
\subfigure[]{
\begin{minipage}[t]{0.32\textwidth}
\centering
\includegraphics[height=3.5cm,width=4.8cm]{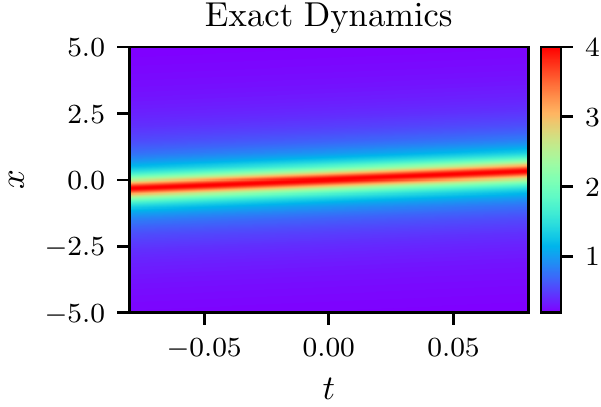}
\end{minipage}
}%
\subfigure[]{
\begin{minipage}[t]{0.32\textwidth}
\centering
\includegraphics[height=3.5cm,width=4.8cm]{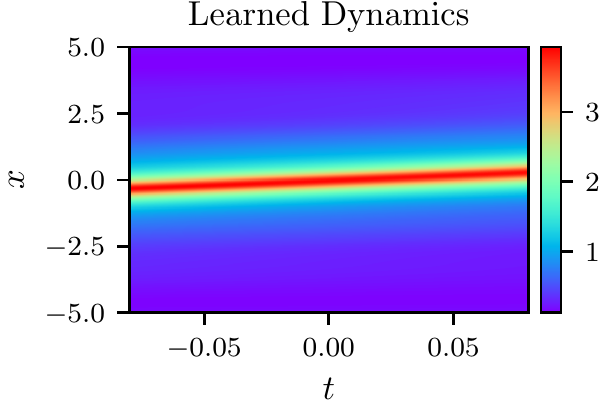}
\end{minipage}%
}%
\subfigure[]{
\begin{minipage}[t]{0.32\textwidth}
\centering
\includegraphics[height=3.5cm,width=4.8cm]{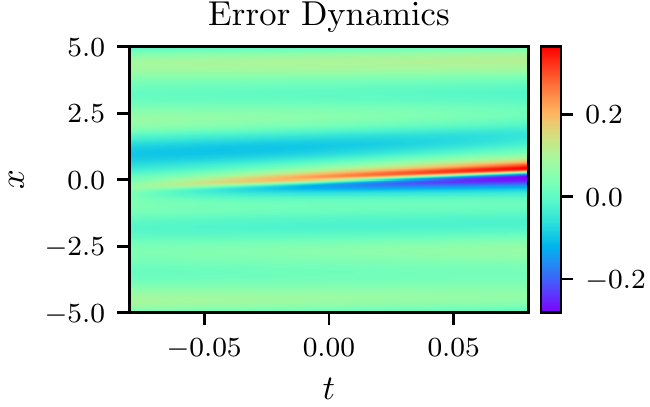}
\end{minipage}%
}%
\centering
\caption{(Color online) The rational solution $q(x,t)$ based on the IPINN: (a) The density plot of exact rational solution; (b) The density plot of learned rational solution; (c) The error density plot of the difference between exact and learned rational solution.}
\label{F2}
\end{figure}

\begin{figure}[htbp]
\centering
\subfigure[]{
\begin{minipage}[t]{0.32\textwidth}
\centering
\includegraphics[height=3.5cm,width=4.8cm]{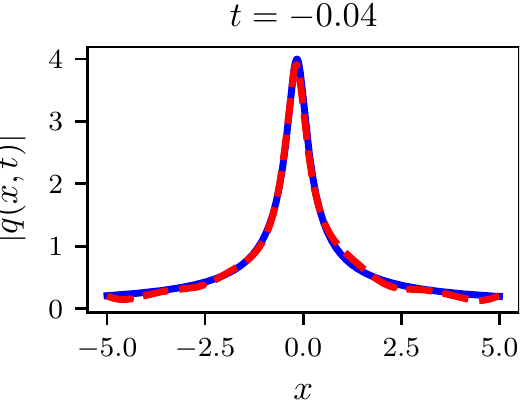}
\end{minipage}
}%
\subfigure[]{
\begin{minipage}[t]{0.32\textwidth}
\centering
\includegraphics[height=3.5cm,width=4.8cm]{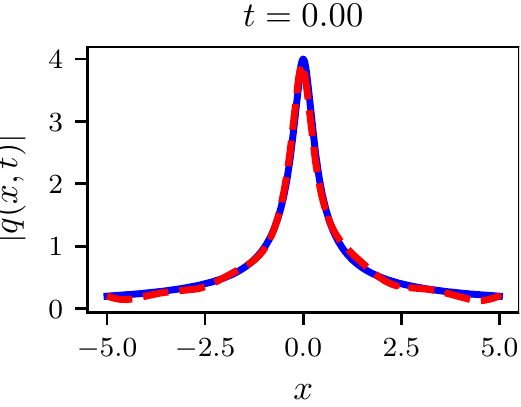}
\end{minipage}%
}%
\subfigure[]{
\begin{minipage}[t]{0.32\textwidth}
\centering
\includegraphics[height=3.5cm,width=4.8cm]{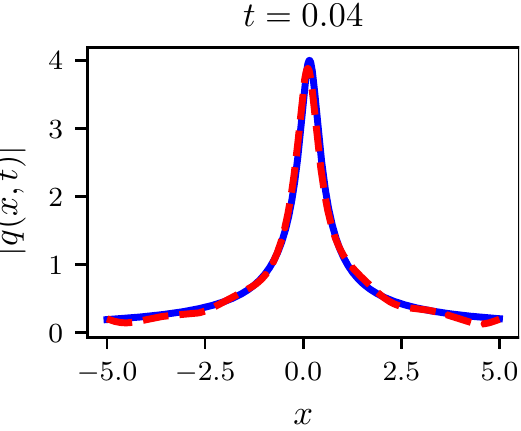}
\end{minipage}%
}%
\centering
\caption{(Color online) The sectional drawings of rational solution $q(x,t)$ based on the IPINN at (a): $t=-0.04$, (b): $t=0$ and (c): $t=0.04$.}
\label{F3}
\end{figure}

\begin{figure}[htbp]
\centering
\subfigure[]{
\begin{minipage}[t]{0.45\textwidth}
\centering
\includegraphics[height=5cm,width=6cm]{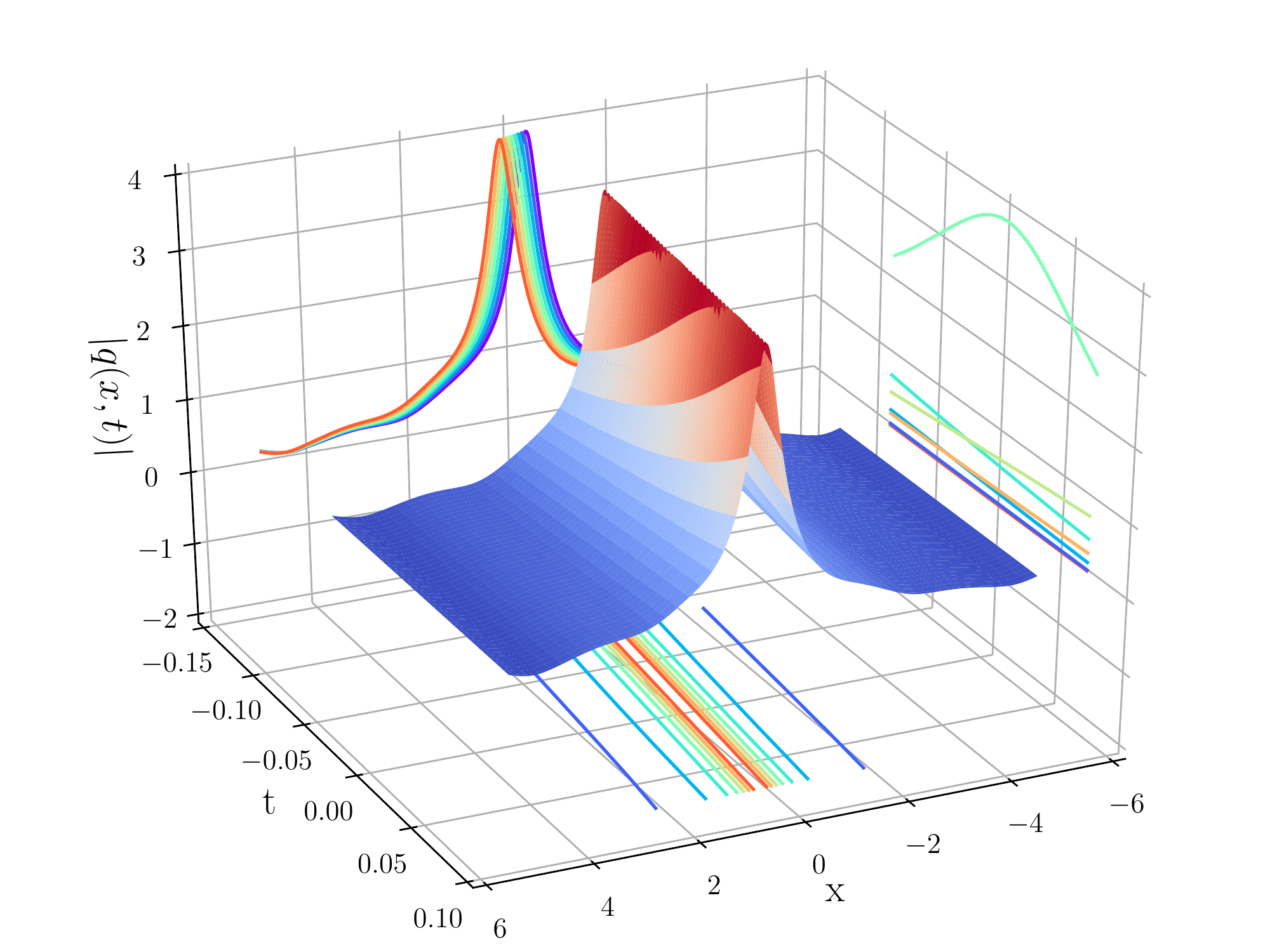}
\end{minipage}
}%
\subfigure[]{
\begin{minipage}[t]{0.45\textwidth}
\centering
\includegraphics[height=5cm,width=6cm]{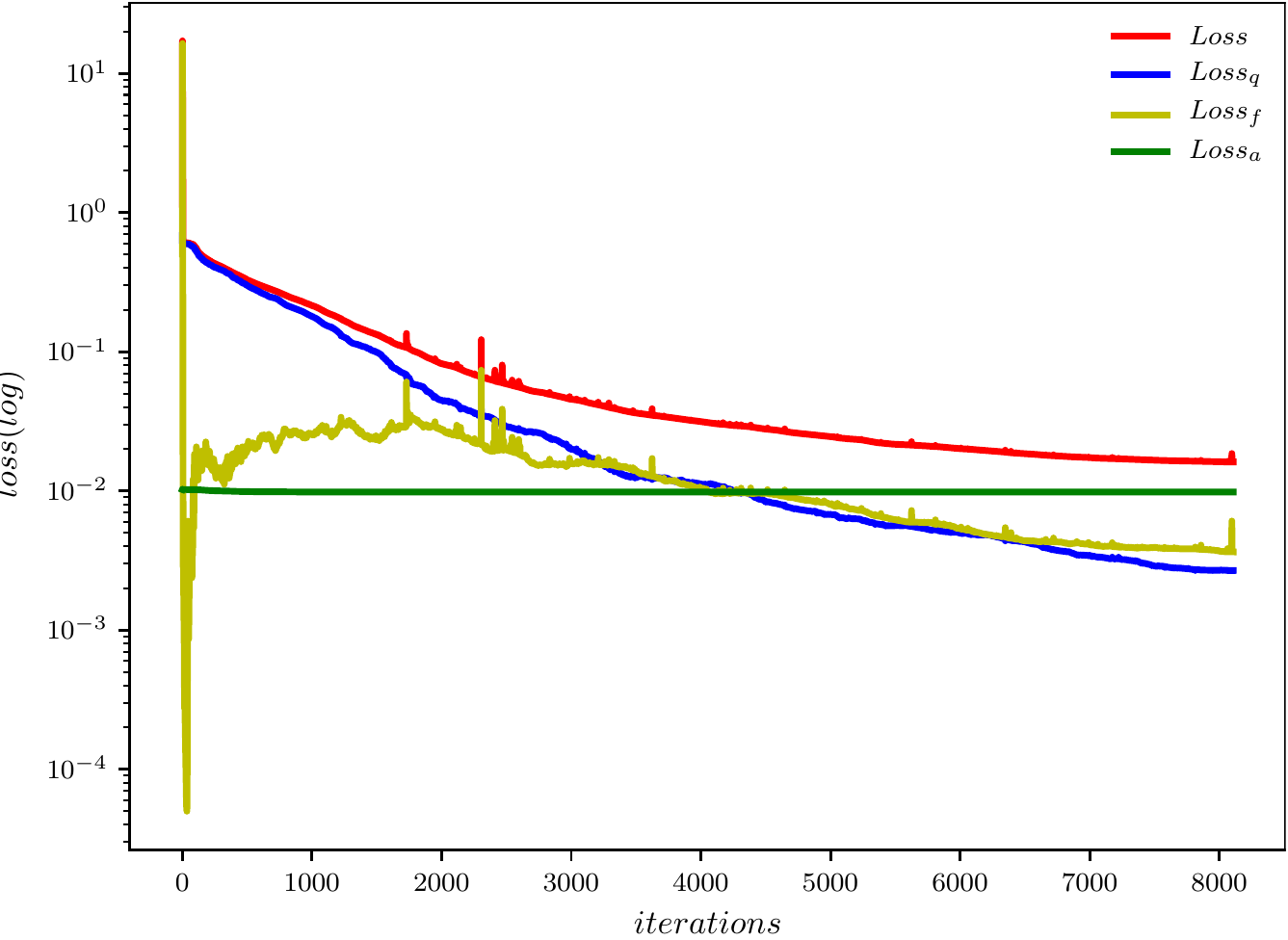}
\end{minipage}
}%
\centering
\caption{(Color online) The rational solution $q(x,t)$ based on the IPINN: (a) The three-dimensional plot; (b) The loss curve figure.}
\label{F4}
\end{figure}

In Figs. \ref{F2} - \ref{F4}, the density plots, the sectional drawing at different times and the iteration number curve plots for the rational solution $q(x,t)$ under IPINN structure are plotted respectively. Specifically, the density plots of exact dynamics, learned dynamics and error dynamics have exhibited in detail, and the corresponding peak scale is shown on the right side of the density plots in Fig. \ref{F2}. Specially, from Fig. \ref{F2} (c), one can obviously find that the error range is about $-0.2$ to $0.2$. In Fig. \ref{F3}, we provide the sectional drawings of rational solution $q(x,t)$ based on the IPINN at (a): $t=-0.04$, (b): $t=0$ and (c): $t=0.04$, and infer that the rational solution propagates right along the $x$-axis. The three-dimensional plot and its corresponding contour map of rational solution for the DNLS \eqref{E1} has been given out in the left panel (a) of Fig. \ref{F4}. From the right panel (a) of Fig. \ref{F4}, we can observe that the $Loss$ curve (red solid line) and the $Loss_q$ curve (blue solid line) converge smoothly, and the $Loss_q$ curve decreases faster. However, the $Loss_f$ curve (yellow solid line) fluctuates greatly and has poor stability. On the contrary, $Loss_a$ curve (green solid line) decreases slowly around $0.01$ and has a strong stability.

\subsection{The data-driven soliton solution}
The explicit soliton solution has been obtained in Ref. \cite{Xu2019}, one can be writen as belows
\begin{align}\label{E15}
q_{\mathrm{ss}}(x,t)=\frac{2\mathrm{i}\left[-\frac12\mathrm{i}\mathrm{cosh}(-6t+2x)+\mathrm{sinh}(-6t+2x)\right]^3\mathrm{e}^{2\mathrm{i}\left(-\frac78t-\frac34x\right)}}{\left\{-\frac54[\mathrm{cosh}(-6t+2x)]^2+1\right\}^2}.
\end{align}

Compared with the nonlinear Schr\"odinger equation (NLS) \cite{Pu2021},  the form of the soliton solution for the DNLS is more complex, and it is more difficult to recover the soliton solution by NNs. Similarly, considering the initial condition $q_{\mathrm{ss}}(x,T_0)$ and Dirichlet boundary conditions $q_{\mathrm{ss}}(L_0,t)$ and $q_{\mathrm{ss}}(L_1,t)$ of Eq. \eqref{E1} arising from the soliton solution Eq. \eqref{E15}, we take $[L_0,L_1]$ and $[T_0,T_1]$ in Eq. \eqref{E1} as $[-3.0,3.0]$ and $[-0.1,0.1]$, respectively. After that, the corresponding Cauchy problem with initial condition $q_0(x)$ can be written as belows
\begin{align}\label{E16}
q_{\mathrm{ss}}(x,-0.1)=\frac{2\mathrm{i}\left[-\frac12\mathrm{i}\mathrm{cosh}(0.6+2x)+\mathrm{sinh}(0.6+2x)\right]^3\mathrm{e}^{2\mathrm{i}\left(0.0875-\frac34x\right)}}{\left\{-\frac54[\mathrm{cosh}(0.6+2x)]^2+1\right\}^2},\,x\in[-3.0,3.0],
\end{align}
and the Dirichlet boundary conditions in Eq. \eqref{E1} become
\begin{align}\label{E17}
q_{\mathrm{lb}}(t)=q_{\mathrm{ss}}(-3.0,t),\,q_{\mathrm{ub}}(t)=q_{\mathrm{ss}}(3.0,t),\,t\in[-0.1,0.1].
\end{align}

In MATLAB, we discretize the Eq. \eqref{E15} by applying the traditional finite difference scheme on even grids, and obtain the training data which contains initial data \eqref{E16} and boundary data \eqref{E17} by dividing the spatial region $[-3.0,3.0]$ into 513 points and the temporal region $[-0.1,0.1]$ into 401 points. We generate a smaller training dataset containing initial-boundary data by randomly extracting $N_q=400$ from original dataset and $N_f=20000$ collocation points which are generated by employing LHS. After giving a dataset of initial and boundary points, the latent soliton solution $q(x,t)$ has been successfully learned by tuning all learnable parameters of the IPINN and regulating the loss function \eqref{E8}. The model of IPINN achieves a relative $\mathbb{L}_2$ error of 5.745440$\mathrm{e}-$02 in about $1417.7192$ seconds, and the number of iterations is 8472.

\begin{figure}[htbp]
\centering
\subfigure[]{
\begin{minipage}[t]{0.32\textwidth}
\centering
\includegraphics[height=3.5cm,width=4.8cm]{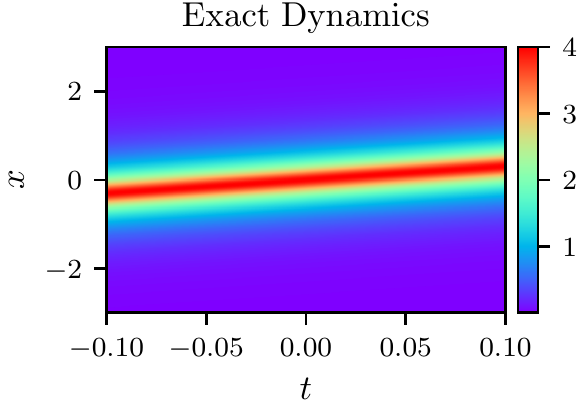}
\end{minipage}
}%
\subfigure[]{
\begin{minipage}[t]{0.32\textwidth}
\centering
\includegraphics[height=3.5cm,width=4.8cm]{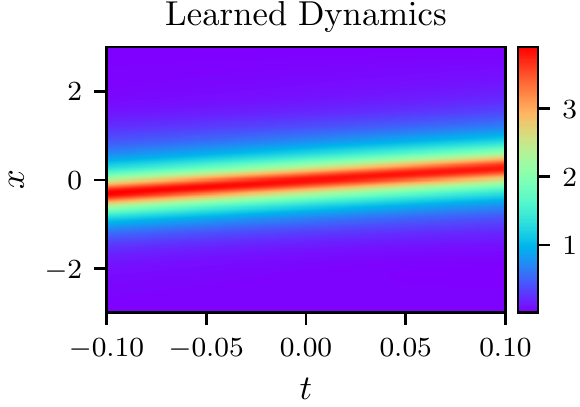}
\end{minipage}%
}%
\subfigure[]{
\begin{minipage}[t]{0.32\textwidth}
\centering
\includegraphics[height=3.5cm,width=4.8cm]{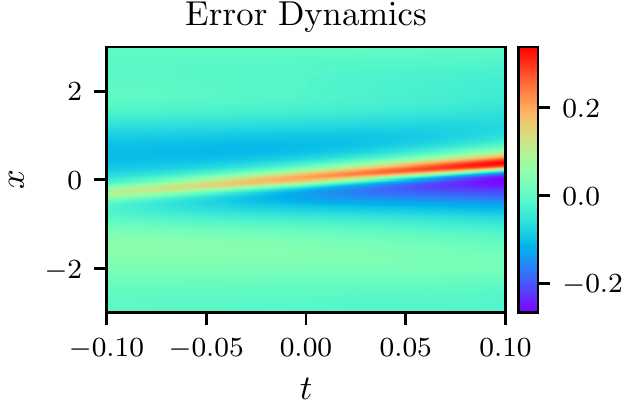}
\end{minipage}%
}%
\centering
\caption{(Color online) The soliton solution $q(x,t)$ based on the IPINN: (a) The density plot of exact soliton solution; (b) The density plot of learned soliton solution; (c) The error density plot of the difference between exact and learned soliton solution.}
\label{F5}
\end{figure}

\begin{figure}[htbp]
\centering
\subfigure[]{
\begin{minipage}[t]{0.32\textwidth}
\centering
\includegraphics[height=3.5cm,width=4.8cm]{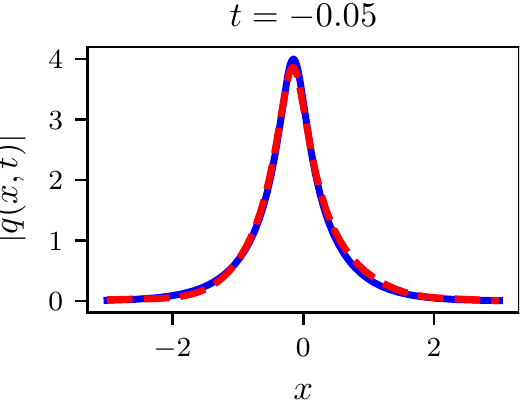}
\end{minipage}
}%
\subfigure[]{
\begin{minipage}[t]{0.32\textwidth}
\centering
\includegraphics[height=3.5cm,width=4.8cm]{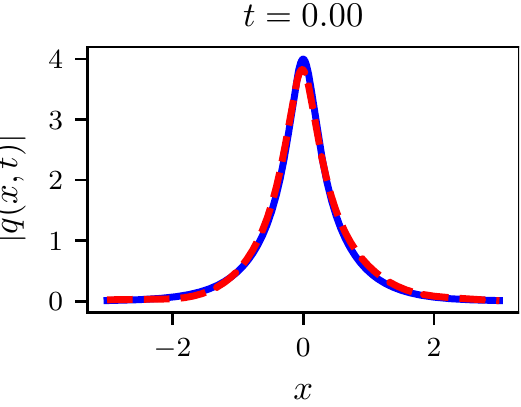}
\end{minipage}%
}%
\subfigure[]{
\begin{minipage}[t]{0.32\textwidth}
\centering
\includegraphics[height=3.5cm,width=4.8cm]{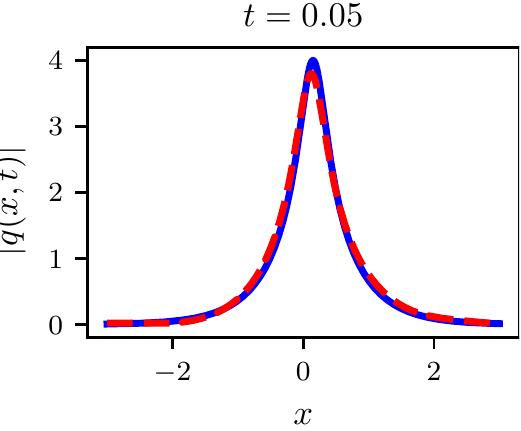}
\end{minipage}%
}%
\centering
\caption{(Color online) The sectional drawings of soliton solution $q(x,t)$ based on the IPINN at (a): $t=-0.05$, (b): $t=0$ and (c): $t=0.05$.}
\label{F6}
\end{figure}

\begin{figure}[htbp]
\centering
\subfigure[]{
\begin{minipage}[t]{0.45\textwidth}
\centering
\includegraphics[height=5cm,width=6cm]{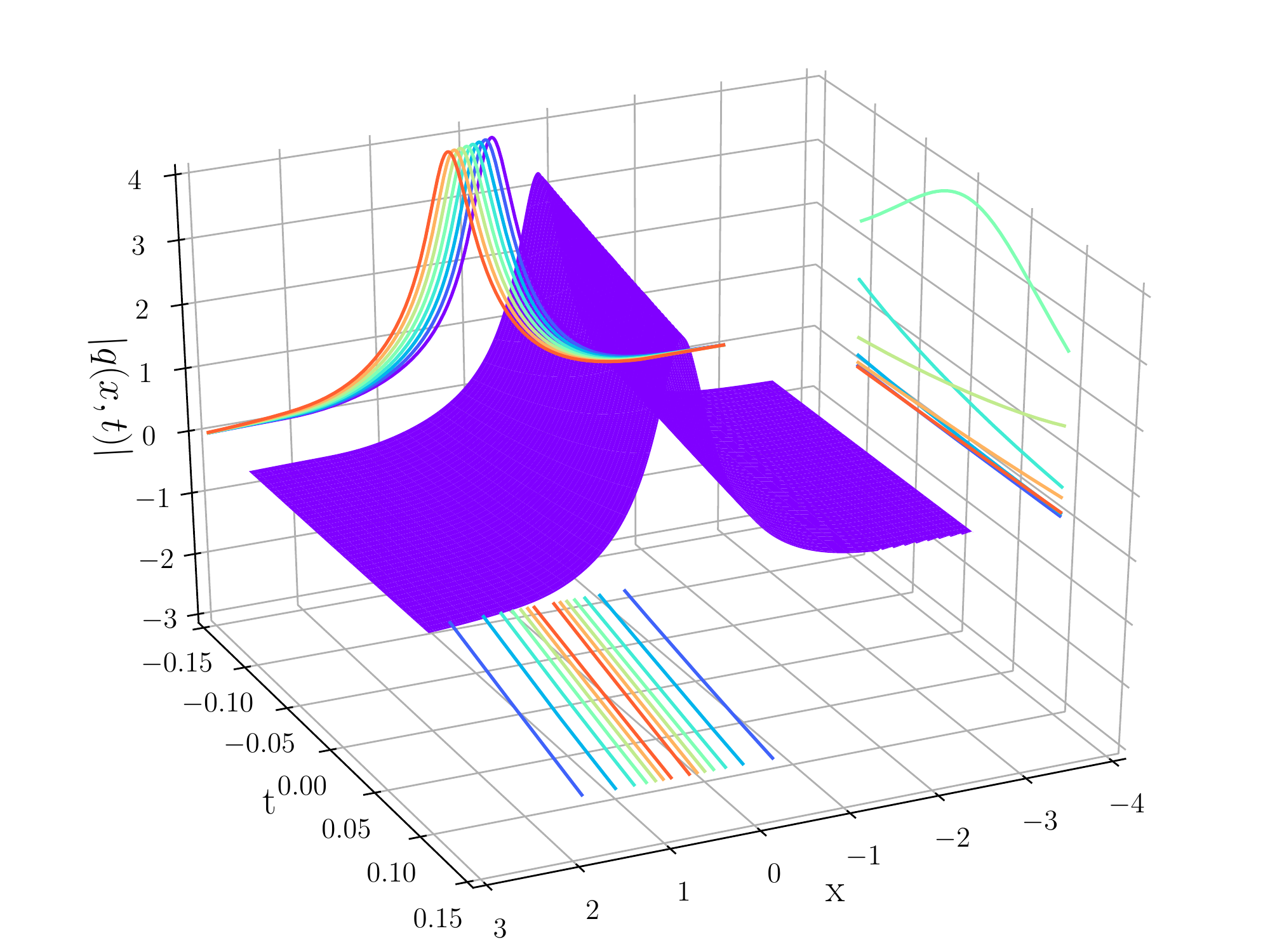}
\end{minipage}
}%
\subfigure[]{
\begin{minipage}[t]{0.45\textwidth}
\centering
\includegraphics[height=5cm,width=6cm]{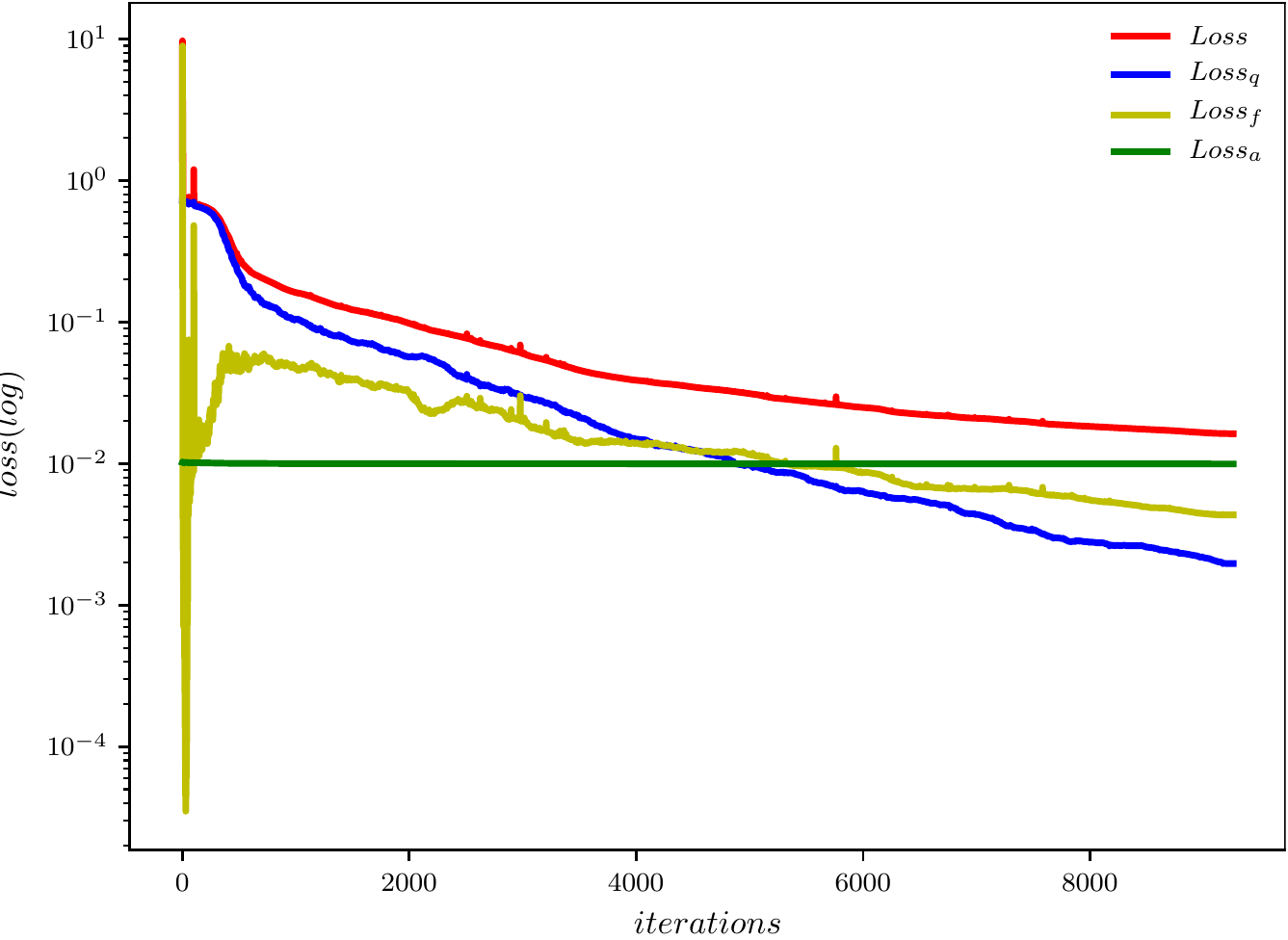}
\end{minipage}
}%
\centering
\caption{(Color online) The soliton solution $q(x,t)$ based on the IPINN: (a) The three-dimensional plot; (b) The loss curve figure.}
\label{F7}
\end{figure}

In Figs. \ref{F5} - \ref{F7}, the density plots, the sectional drawing at different times and the iteration number curve plots for the soliton solution $q(x,t)$ under IPINN structure are plotted respectively. Specifically, the density plots of exact dynamics, learned dynamics and error dynamics have exhibited in detail, and the corresponding peak scale is shown on the right side of the density plots in Fig. \ref{F5}. Similar to Fig. \ref{F2} (c), it is obvious that the error range is also about $-0.2$ to $0.2$ from Fig. \ref{F5} (c). In Fig. \ref{F6}, the sectional drawings with different time point for soliton solution $q(x,t)$ have been given out respectively. Similarly with Section 3.1, when the increase of time $t$, the soliton solution propagates from left to right along the $x$-axis. The left panel (a) of Fig. \ref{F7} exhibits the three-dimensional plot and its corresponding contour map of soliton solution for the DNLS \eqref{E1}. From the right panel (b) of Fig. \ref{F7}, one can observe that the $Loss$ curve (red solid line) and the $Loss_q$ curve (blue solid line) converge smoothly, and the $Loss_q$ curve decreases faster. On the other hand, the $Loss_f$ curve (yellow solid line) fluctuates greatly and has poor stability. Similarly, $Loss_a$ curve (green solid line) decreases slowly around $0.01$ and has a strong stability.

Comparing the rational solution with the soliton solution, we can find that the two solutions have different forms, in which the initial and boundary conditions of the rational solution is simpler, but the shapes of the two solutions are similar. Through a large number of numerical training experiments, we find that it is more difficult to simulate the ratinal solution and the soliton solution for the DNLS than the NLS equation, and the range of time $t$ should not be too large, otherwise the training error will become large.

\section{The data-driven periodic wave, rogue wave and rogue periodic wave of the DNLS}

It is well known that periodic waves and rogue waves are very significant waves in nonlinear integrable systems, and the periodic wave and rogue wave of the DNLS have been also widely studied \cite{Steudel2003,Guo2012,Xue2020,Xu2019,Xu2011,Yang2020,ZhangY2014,Liu2018,ChenJB2021,PuJ2021}. However, to the best of our knowledge, the mixtures of periodic solutions and rogue-wave solutions for the DNLS were not considered up to now by utilizing deep learning method based on NNs. Thus, it is interesting and necessary to find the data-driven rogue periodic wave solution describing the behavior of rogue waves on a periodic background. In Ref. \cite{PuJ2021}, Pu et al. have investigated the rational solutions and two order rogue wave solution for another form of the DNLS by classical PINN and IPINN, and pointed out the IPINN has more advantages about the overall effect, especially in simulation of the complex rogue wave solutions. Furthermore, as far as we know, the periodic wave and one order rogue wave of the DNLS have also not been investigated yet by employing PINNs technique. Therefore, in this section, we will devote to investigate the data-driven periodic wave, rogue wave and rogue periodic wave of the DNLS with the aid of IPINN, which consists of nine hidden layers and each layer has 40 neurons.

\subsection{The data-driven periodic wave solution}
From Ref. \cite{Liu2018}, the explicit periodic wave solution of the DNLS can be obtained as
\begin{align}\label{E18}
q_{\mathrm{pw}}(x,t)=\frac{\left[(\sqrt{5}+2)\mathrm{e}^{\frac{\sqrt{5}}{2}\mathrm{i}x}+\left(-\frac12-\frac{\sqrt{5}}{2}\right)\mathrm{e}^{\frac{\sqrt{5}}{2}\mathrm{i}(t-x)}+\left(-\frac32-\frac{\sqrt{5}}{2}\right)\mathrm{e}^{\frac{\sqrt{5}}{4}\mathrm{i}t}\right]\mathrm{e}^{-\mathrm{i}x}}{\left[\mathrm{e}^{\frac{\sqrt{5}}{4}\mathrm{i}x}+\left(\frac32+\frac{\sqrt{5}}{2}\right)\mathrm{e}^{\frac{\sqrt{5}}{4}\mathrm{i}(t-x)}\right]^2}.
\end{align}

Eq. \eqref{E18} is a periodic wave solution which is periodic in space $x$, and it is obvious that the period is 3 when $x\in[-6,12]$ , the amplitude of each wave crest and trough of wave is 2 and 0, respectively. Since the waveforms of periodic waves on both sides of time $t=0$ are the same, the range of time $t$ does not need to be symmetric about $t=0$. Therefore, one can take $[L_0,L_1]$ and $[T_0,T_1]$ in Eq. \eqref{E1} as $[-6.0,12.0]$ and $[0.0,2.0]$ in this subsection, respectively. The corresponding Cauchy problem with the initial condition $q_0(x)$ of Eq. \eqref{E1} arising from the periodic wave solution Eq. \eqref{E18} can be written as belows
\begin{align}\label{E19}
q_{\mathrm{pw}}(x,0.0)=\frac{\left[(\sqrt{5}+2)\mathrm{e}^{\frac{\sqrt{5}}{2}\mathrm{i}x}+\left(-\frac12-\frac{\sqrt{5}}{2}\right)\mathrm{e}^{-\frac{\sqrt{5}}{2}\mathrm{i}x}-\frac32-\frac{\sqrt{5}}{2}\right]\mathrm{e}^{-\mathrm{i}x}}{\left[\mathrm{e}^{\frac{\sqrt{5}}{4}\mathrm{i}x}+\left(\frac32+\frac{\sqrt{5}}{2}\right)\mathrm{e}^{-\frac{\sqrt{5}}{4}\mathrm{i}x}\right]^2},\,x\in[-6.0,12.0],
\end{align}
and the Dirichlet boundary conditions $q(L_0,t)$ and $q(L_1,t)$ in Eq. \eqref{E1} become as following
\begin{align}\label{E20}
q_{\mathrm{lb}}(t)=q_{\mathrm{pw}}(-6.0,t),\,q_{\mathrm{ub}}(t)=q_{\mathrm{pw}}(12.0,t),\,t\in[0.0,2.0].
\end{align}

Similarly, discretizing Eq. \eqref{E18} via the aid of the traditional finite difference scheme on even grids, and obtain the original training data which contain initial data \eqref{E19} and boundary data \eqref{E20} by dividing the spatial region $[-6.0,12.0]$ into 513 points and the temporal region $[0.0,2.0]$ into 401 points. Then, one can generate a smaller training dataset that containing initial-boundary data by randomly extracting $N_q=100$ from original dataset and $N_f=10000$ collocation points which are produced by the LHS. After that, the latent periodic wave solution $q(x,t)$ has been successfully learned by tuning all learnable parameters of the IPINN and regulating the loss function \eqref{E8}. The model of IPINN achieves a relative $\mathbb{L}_2$ error of 2.035710$\mathrm{e}-$02 in about 2827.0457 seconds, and the number of iterations is 20157.

\begin{figure}[htbp]
\centering
\subfigure[]{
\begin{minipage}[t]{0.32\textwidth}
\centering
\includegraphics[height=3.5cm,width=4.8cm]{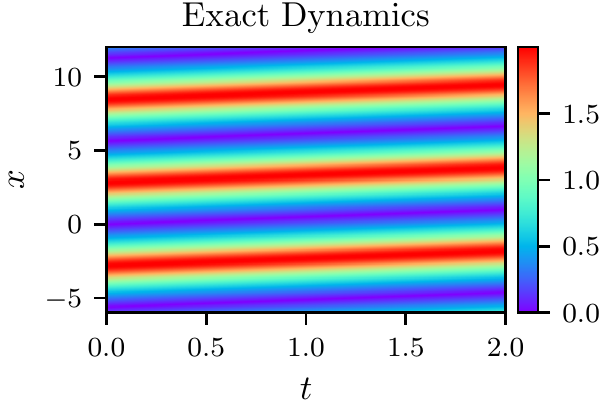}
\end{minipage}
}%
\subfigure[]{
\begin{minipage}[t]{0.32\textwidth}
\centering
\includegraphics[height=3.5cm,width=4.8cm]{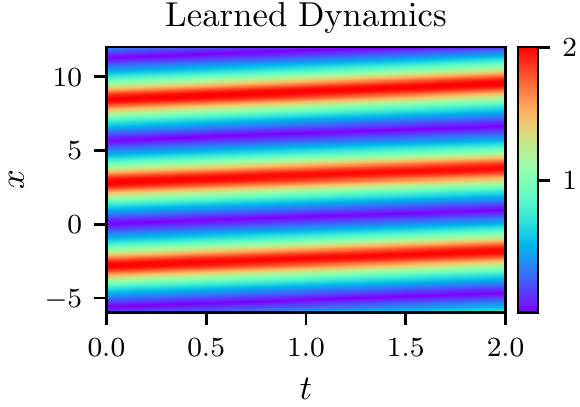}
\end{minipage}%
}%
\subfigure[]{
\begin{minipage}[t]{0.32\textwidth}
\centering
\includegraphics[height=3.5cm,width=4.8cm]{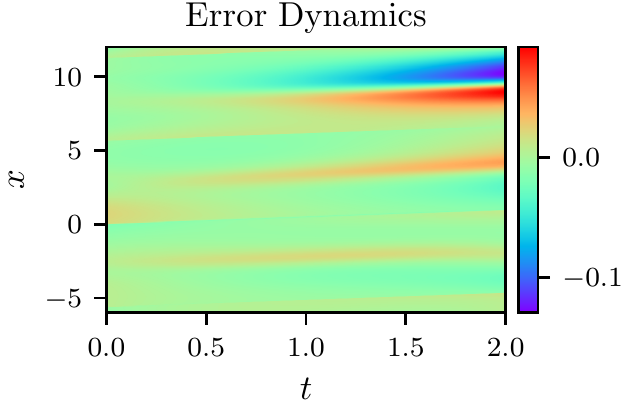}
\end{minipage}%
}%
\centering
\caption{(Color online) The periodic wave solution $q(x,t)$ based on the IPINN: (a) The density plot of exact periodic wave solution; (b) The density plot of learned periodic wave solution; (c) The error density plot of the difference between exact and learned periodic wave solution.}
\label{F8}
\end{figure}

\begin{figure}[htbp]
\centering
\subfigure[]{
\begin{minipage}[t]{0.32\textwidth}
\centering
\includegraphics[height=3.5cm,width=4.8cm]{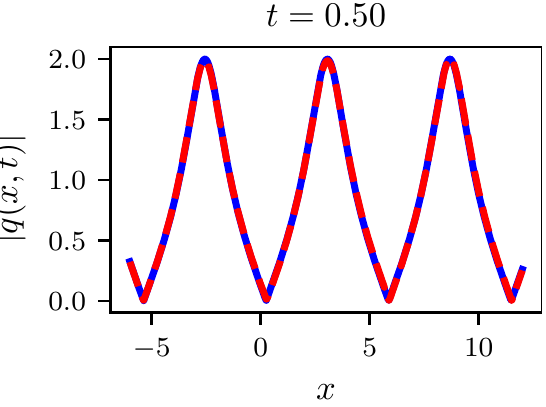}
\end{minipage}
}%
\subfigure[]{
\begin{minipage}[t]{0.32\textwidth}
\centering
\includegraphics[height=3.5cm,width=4.8cm]{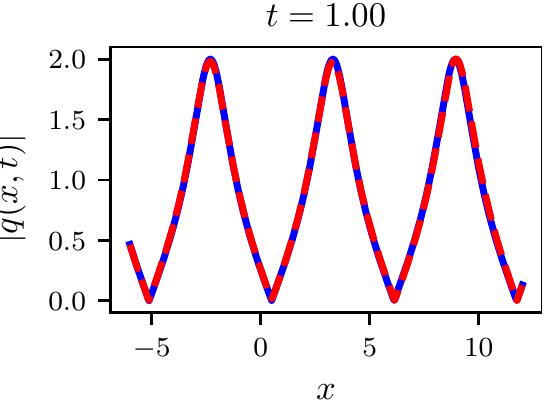}
\end{minipage}%
}%
\subfigure[]{
\begin{minipage}[t]{0.32\textwidth}
\centering
\includegraphics[height=3.5cm,width=4.8cm]{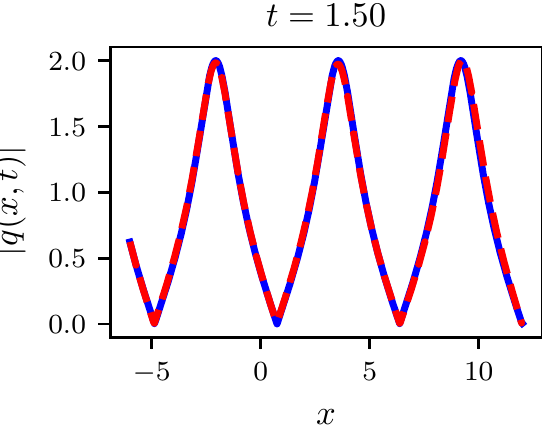}
\end{minipage}%
}%
\centering
\caption{(Color online) The sectional drawings of periodic wave solution $q(x,t)$ based on the IPINN at (a): $t=0.50$, (b): $t=1.00$ and (c): $t=1.50$.}
\label{F9}
\end{figure}

\begin{figure}[htbp]
\centering
\subfigure[]{
\begin{minipage}[t]{0.45\textwidth}
\centering
\includegraphics[height=5cm,width=6cm]{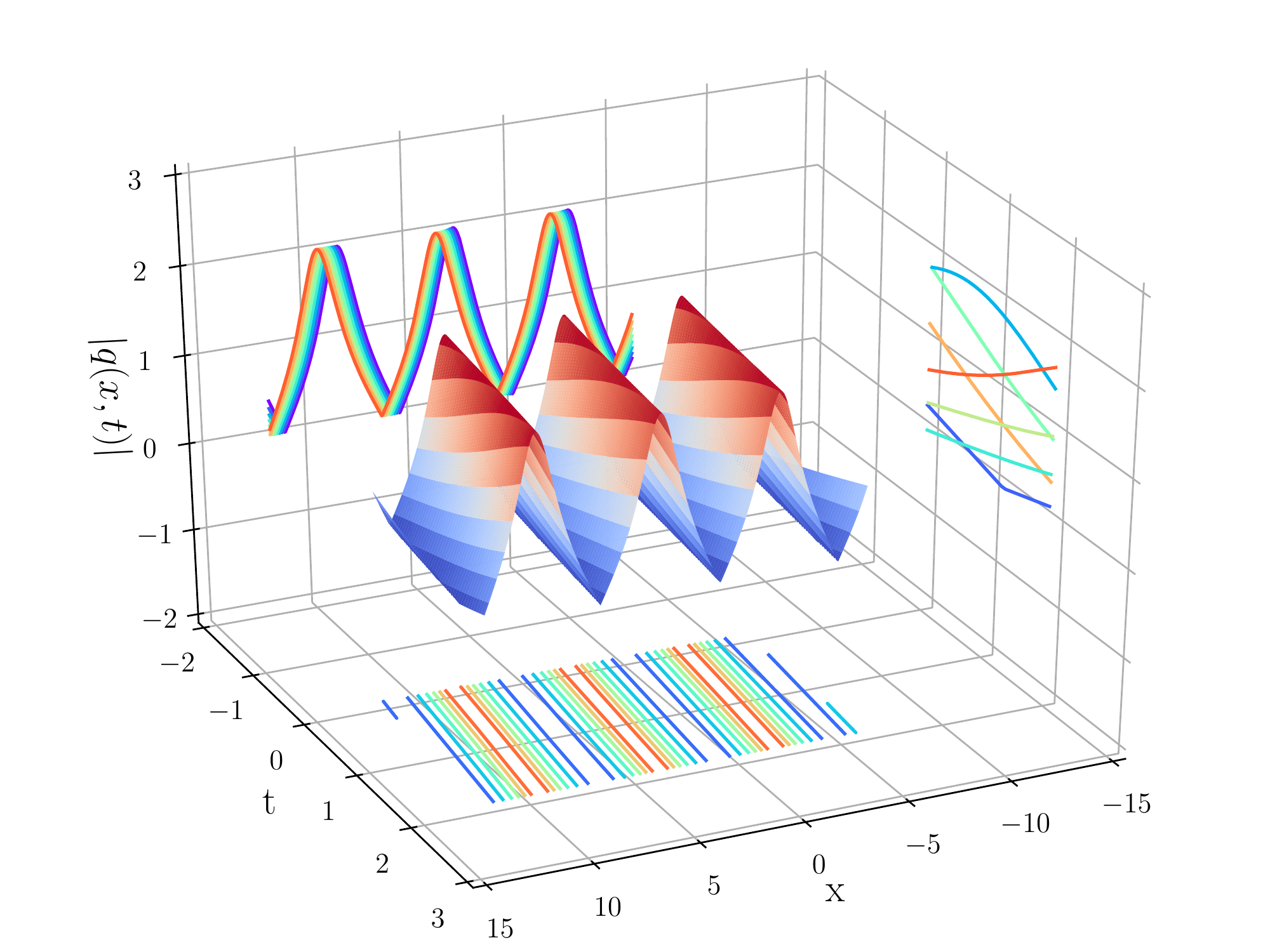}
\end{minipage}
}%
\subfigure[]{
\begin{minipage}[t]{0.45\textwidth}
\centering
\includegraphics[height=5cm,width=6cm]{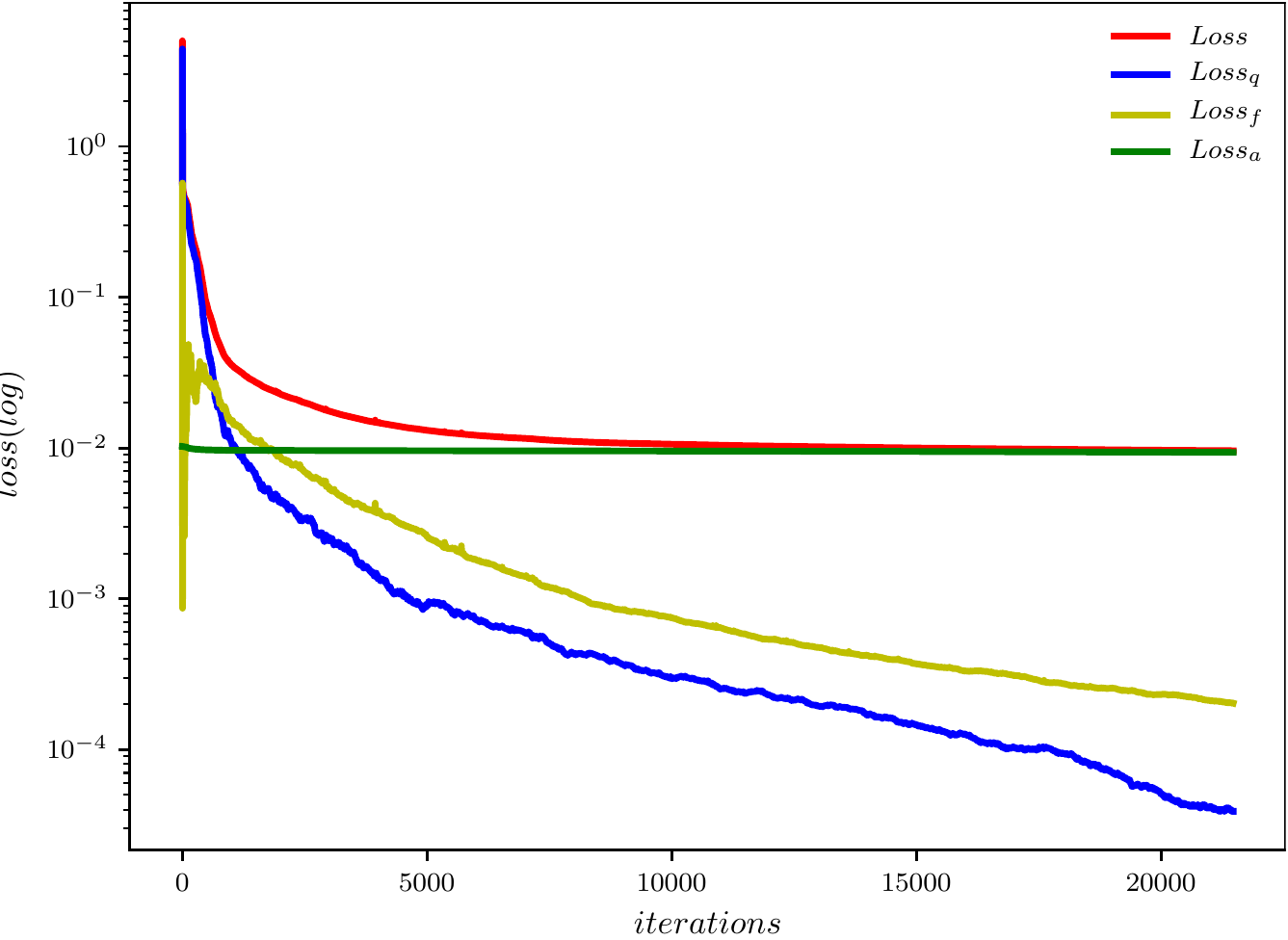}
\end{minipage}
}%
\centering
\caption{(Color online) The periodic wave solution $q(x,t)$ based on the IPINN: (a) The three-dimensional plot; (b) The loss curve figure.}
\label{F10}
\end{figure}

In Figs. \ref{F8} - \ref{F10}, the density plots, the sectional drawing at different times and the Loss curve plots for the periodic wave solution $q(x,t)$ under IPINN structure are plotted respectively. Specifically, the density plots of exact dynamics, learned dynamics and error dynamics have exhibited in detail, and the corresponding peak scale is shown on the right side of the density plots in Fig. \ref{F8}. From the striped density map, it is obvious that the periodic wave has three periods and the amplitude is the same. Specially, according to the Fig. \ref{F8} (c), one can obviously find that the error range is about $-0.1$ to $0.1$, this error range is lower than that of the two data-driven solutions in Section 3. In Fig. \ref{F9}, we provide the sectional drawings of rational solution $q(x,t)$ based on the IPINN at (a): $t=0.50$, (b): $t=1.00$ and (c): $t=1.50$, and infer that the periodic wave solution propagates right along the $x$-axis as time $t$ increases. The three-dimensional plot and its corresponding contour map of periodic wave solution for the DNLS \eqref{E1} has been given out in the left panel (a) of Fig. \ref{F10}. From the right panel (b) of Fig. \ref{F10}, we can observe that the $Loss$ curve (red solid line) and $Loss_q$ curve (blue solid line) converge smoothly, in which the $Loss_q$ curve decreases faster. However, the $Loss_f$ curve (yellow solid line) fluctuates violently in the first 100 iterations and converges smoothly in the later iterations, the convergence rate of $Loss_f$ is between the $Loss$ curve and the $Loss_q$ curve. Here, $Loss_a$ curve (green solid line) decreases slowly around $0.01$ and has a strong stability.

\subsection{The data-driven rogue wave solution}
From Ref. \cite{ZhangY2014}, the rogue wave solution of the DNLS can be derived as
\begin{align}\label{E21}
q_{\mathrm{rw}}(x,t)=\frac{[-2t^2-2x^2-1-\mathrm{i}(-2t+2x)][2t^2+2x^2-3+\mathrm{i}(6t+2x)]\mathrm{e}^{\mathrm{i}x}}{[-2t^2-2x^2-1+\mathrm{i}(-2t+2x)]^2}.
\end{align}

Similarly, considering the initial condition $q_{\mathrm{rw}}(x,T_0)$ and Dirichlet boundary condition $q_{\mathrm{rw}}(L_0,t)$ and $q_{\mathrm{rw}}(L_1,t)$ of Eq. \eqref{E1} arising from the rogue wave solution Eq. \eqref{E21}, the $[L_0,L_1]$ and $[T_0,T_1]$ in Eq. \eqref{E1} are taken as $[-10.0,10.0]$ and $[-1.0,1.0]$, respectively. After that, the corresponding the Cauchy problem with initial condition $q_0(x)$ can be written as belows
\begin{align}\label{E22}
q_{\mathrm{rw}}(x,-1.0)=\frac{[-2-2x^2-1-\mathrm{i}(2+2x)][2+2x^2-3+\mathrm{i}(-6+2x)]\mathrm{e}^{\mathrm{i}x}}{[-2-2x^2-1+\mathrm{i}(2+2x)]^2},\,x\in[-10.0,10.0],
\end{align}
and the Dirichlet boundary condition evolve into
\begin{align}\label{E23}
q_{\mathrm{lb}}(t)=q_{\mathrm{rw}}(-10.0,t),\,q_{\mathrm{ub}}(t)=q_{\mathrm{rw}}(10.0,t),\,t\in[-1.0,1.0].
\end{align}

Similar to Section 4.1, we discretize the Eq. \eqref{E21} by applying the traditional finite difference scheme on even grids with the help of Matlab, and obtain the training data which contain initial data \eqref{E22} and boundary data \eqref{E23} by dividing the spatial region $[-10.0,10.0]$ into 513 points and the temporal region $[-1.0,1.0]$ into 401 points. A smaller training dataset containing initial-boundary data will be generated by randomly extracting $N_q=400$ from original dataset and $N_f=20000$ collocation points via LHS. After giving a dataset of initial and boundary points, the latent rogue wave solution $q(x,t)$ has been successfully learned by tuning all learnable parameters of the IPINN and regulating the loss function \eqref{E8}. The model of IPINN achieves a relative $\mathbb{L}_2$ error of 7.459070$\mathrm{e}-$02 in about 2833.9749 seconds, and the number of iterations is 18138.

\begin{figure}[htbp]
\centering
\subfigure[]{
\begin{minipage}[t]{0.32\textwidth}
\centering
\includegraphics[height=3.5cm,width=4.8cm]{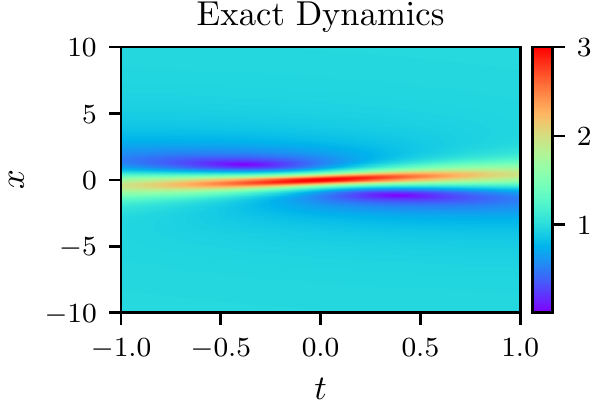}
\end{minipage}
}%
\subfigure[]{
\begin{minipage}[t]{0.32\textwidth}
\centering
\includegraphics[height=3.5cm,width=4.8cm]{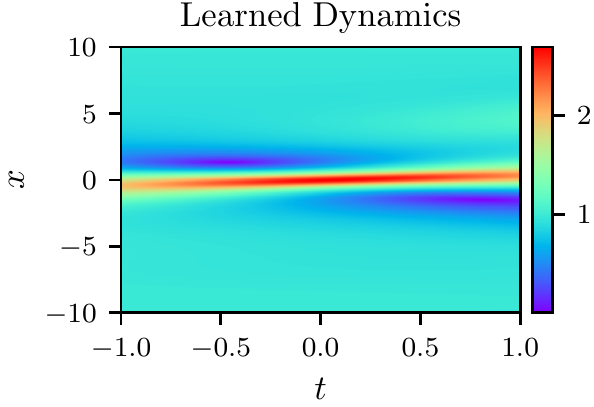}
\end{minipage}%
}%
\subfigure[]{
\begin{minipage}[t]{0.32\textwidth}
\centering
\includegraphics[height=3.5cm,width=4.8cm]{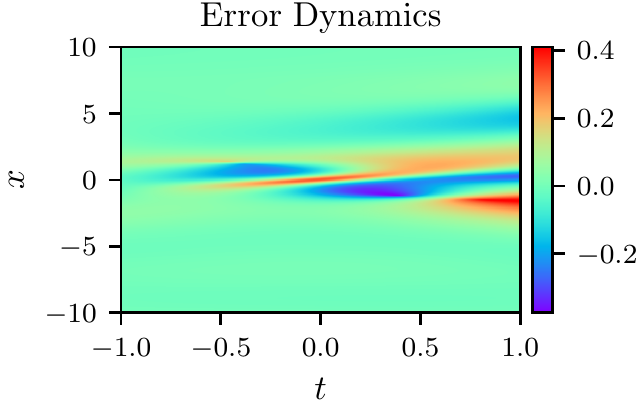}
\end{minipage}%
}%
\centering
\caption{(Color online) The rogue wave solution $q(x,t)$ based on the IPINN: (a) The density plot of exact rogue wave solution; (b) The density plot of learned rogue wave solution; (c) The error density plot of the difference between exact and learned rogue wave solution.}
\label{F11}
\end{figure}

\begin{figure}[htbp]
\centering
\subfigure[]{
\begin{minipage}[t]{0.32\textwidth}
\centering
\includegraphics[height=3.5cm,width=4.8cm]{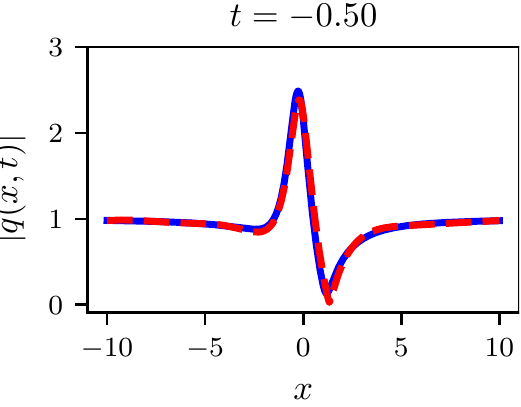}
\end{minipage}
}%
\subfigure[]{
\begin{minipage}[t]{0.32\textwidth}
\centering
\includegraphics[height=3.5cm,width=4.8cm]{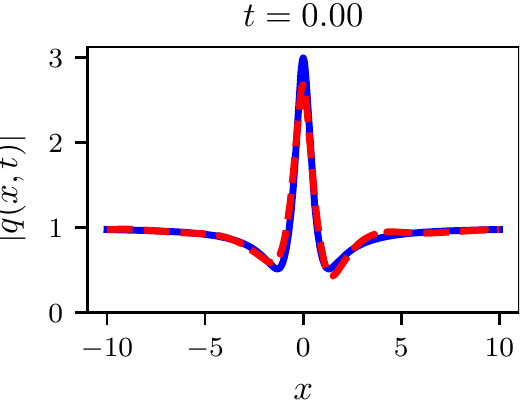}
\end{minipage}%
}%
\subfigure[]{
\begin{minipage}[t]{0.32\textwidth}
\centering
\includegraphics[height=3.5cm,width=4.8cm]{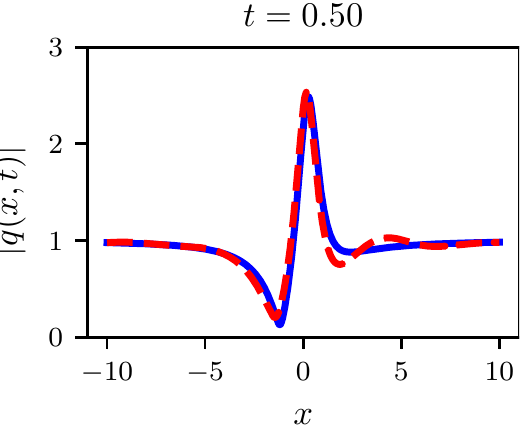}
\end{minipage}%
}%
\centering
\caption{(Color online) The sectional drawings of rogue wave solution $q(x,t)$ based on the IPINN at (a): $t=-0.5$, (b): $t=0$ and (c): $t=0.5$.}
\label{F12}
\end{figure}

\begin{figure}[htbp]
\centering
\subfigure[]{
\begin{minipage}[t]{0.45\textwidth}
\centering
\includegraphics[height=5cm,width=6cm]{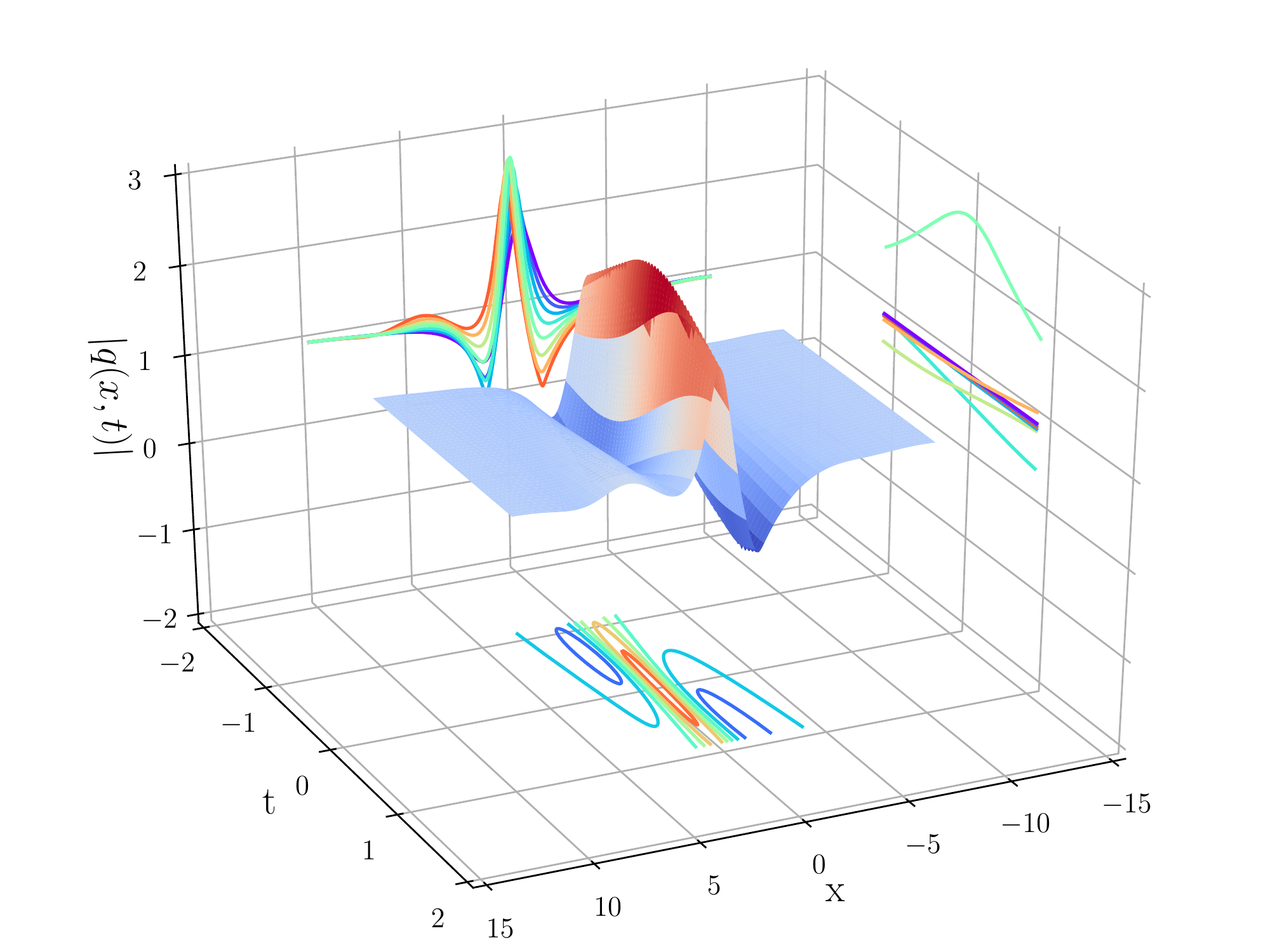}
\end{minipage}
}%
\subfigure[]{
\begin{minipage}[t]{0.45\textwidth}
\centering
\includegraphics[height=5cm,width=6cm]{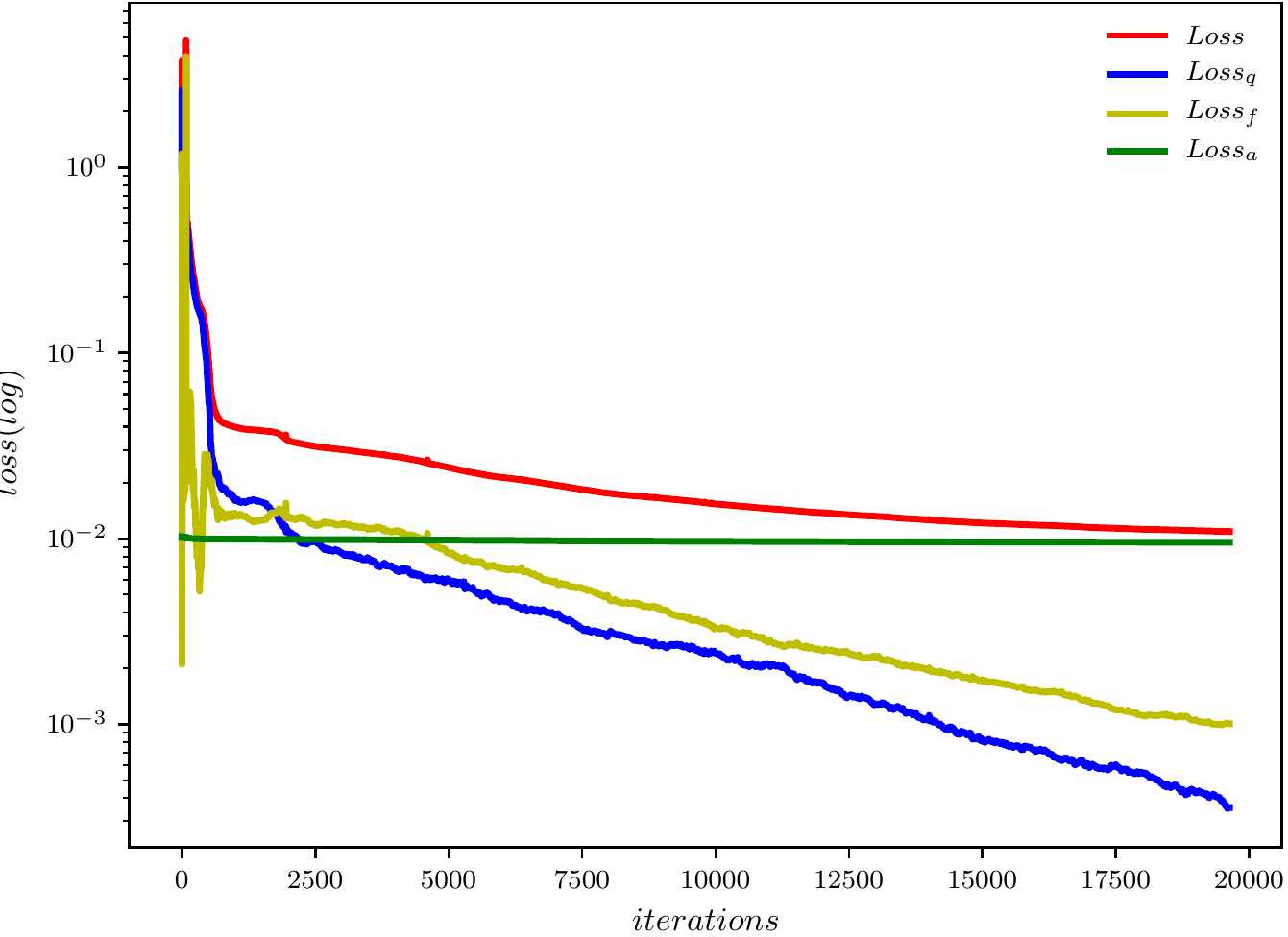}
\end{minipage}
}%
\centering
\caption{(Color online) The rogue wave solution $q(x,t)$ based on the IPINN: (a) The three-dimensional plot; (b) The loss curve figure.}
\label{F13}
\end{figure}

In Figs. \ref{F11} - \ref{F13}, the density plots, the sectional drawing at different times and the iteration number curve plots for the rogue wave solution $q(x,t)$ under IPINN structure are plotted respectively. Specifically, the density plots with the corresponding peak scale for exact dynamics, learned dynamics and error dynamics have exhibited in Fig. \ref{F11}. Specially, by analysing the Fig. \ref{F11} (c), one can obviously find that the error range is about $-0.4$ to $0.4$, the error range is obviously larger than that in Section 3. In Fig. \ref{F12}, the sectional drawings for rogue wave solution $q(x,t)$ based on the IPINN at (a): $t=-0.50$, (b): $t=0$ and (c): $t=0.50$ have been provide, and infering that the amplitude of the rogue wave solution reaches the maximum at time $t=0$, and the waveform of the profiles at time $t=-0.50$ and $t=0.50$ are symmetrical. The three-dimensional plot with corresponding contour map of rogue wave solution for the DNLS \eqref{E1} has been given out in the left panel (a) of Fig. \ref{F13}. From the right panel (b) of Fig. \ref{F13}, we can observe that the $Loss$ curve (red solid line) converge smoothly, and and the $Loss_q$ curve (blue solid line) decreases faster. However, when the number of iterations is less than 2500, the $Loss_f$ curve (yellow solid line) fluctuates greatly and has poor stability. Once the number of iterations is greater than 2500, the descent speed of $Loss_f$ curve is almost the same as that of $Loss_q$ curve. The same as before, $Loss_a$ curve (green solid line) decreases slowly around $0.01$ and has a strong stability.

\subsection{The data-driven rogue periodic wave solution}
Compared with the traditional rogue wave which is on plane wave background, the rogue periodic wave is the rogue wave solution on the periodic wave background. From Ref. \cite{ZhangY2014}, the rogue periodic wave solution of the DNLS can be obtained as follows
\begin{align}\label{E24}
q_{\mathrm{rpw}}(x,t)=\frac{\Omega_{11}^2}{\Omega_{21}^2}\mathrm{e}^{\mathrm{i}x}+2\mathrm{i}\frac{\Omega_{11}\Omega_{12}}{\Omega_{21}^2},
\end{align}
where
\begin{align}\nonumber
\begin{split}
\Omega_{11}=&\frac{1}{15625}\Big\{\big[(2498\mathrm{i}t-2498\mathrm{i}x-2502t^2-2502x^2-1251)\sqrt{1562501}+3119998\mathrm{i}t-\\
&3129998\mathrm{i}x-3125002t^2-3125002x^2-1567501\big]\mathrm{e}^{\frac{-\mathrm{i}(t-1250x)\sqrt{1562501}}{3125000}-\frac{\mathrm{i}x}{2}}\Big\}-\frac{4}{625}\\
&\bigg[\bigg(\mathrm{i}t+\mathrm{i}x-t^2-x^2+\frac12\bigg)\sqrt{1562501}+2\mathrm{i}t+2500\mathrm{i}x+1251\bigg]\mathrm{e}^{\frac{\mathrm{i}(t-1250x)\sqrt{1562501}}{3125000}-\frac{\mathrm{i}x}{2}},
\end{split}
\end{align}
\begin{align}\nonumber
\begin{split}
\Omega_{12}=&\frac{1}{781250}\Big\{\big[(-125000\mathrm{i}t^2+62500\mathrm{i}-125000\mathrm{i}x^2+125000t+125000x)\sqrt{1562501}-\\
&156250100\mathrm{i}(t^2+x^2)+78249950\mathrm{i}+156500100t+156249900x\big]\\
&\mathrm{e}^{\frac{-\mathrm{i}(t-1250x)\sqrt{1562501}}{3125000}+\frac{\mathrm{i}x}{2}}\Big\}+\frac{1}{390625}\bigg\{1249\bigg[\bigg(\mathrm{i}t^2+\mathrm{i}x^2+\frac{\mathrm{i}}{2}+\frac{1251t}{1249}-\frac{1251x}{1249}\bigg)\sqrt{1562501}\\
&-\frac{1562501\mathrm{i}}{1249}(t^2+x^2)+\frac{4687499\mathrm{i}}{2498}+\frac{4690001t}{1249}+\frac{1559999x}{1249}\bigg]\mathrm{e}^{\frac{\mathrm{i}(t-1250x)\sqrt{1562501}}{3125000}+\frac{\mathrm{i}x}{2}}\bigg\},
\end{split}
\end{align}
and
\begin{align}\nonumber
\begin{split}
\Omega_{21}=&\frac{1}{15625}\Big\{\big[(-2498\mathrm{i}t-2502t^2-2502x^2+2498\mathrm{i}x-1251)\sqrt{1562501}-3119998\mathrm{i}t+\\
&3129998\mathrm{i}x-3125002(t^2+x^2)-1567501\big]\mathrm{e}^{\frac{\mathrm{i}(t-1250x)\sqrt{1562501}}{3125000}+\frac{\mathrm{i}x}{2}}\Big\}+\frac{4}{625}\bigg[\bigg(\mathrm{i}t+\mathrm{i}x+t^2\\
&+x^2-\frac12\bigg)\sqrt{1562501}+2\mathrm{i}t+2500\mathrm{i}x-1251\bigg]\mathrm{e}^{\frac{-\mathrm{i}(t-1250x)\sqrt{1562501}}{3125000}+\frac{\mathrm{i}x}{2}}.
\end{split}
\end{align}

Similarly, considering the initial condition $q_{\mathrm{rpw}}(x,T_0)$ and Dirichlet boundary conditions $q_{\mathrm{rpw}}(L_0,t)$ and $q_{\mathrm{rpw}}(L_1,t)$ of Eq. \eqref{E1} arising from the rogue periodic wave solution Eq. \eqref{E24}, we set $[L_0,L_1]$ and $[T_0,T_1]$ in Eq. \eqref{E1} as $[-20.0,20.0]$ and $[-0.5,0.5]$, respectively. After that, the corresponding the Cauchy problem with initial condition $q_0(x)$ can be written as belows
\begin{align}\label{E25}
q_{\mathrm{rpw}}(x,-0.5)=\frac{\Omega_{11}'^2}{\Omega_{21}'^2}\mathrm{e}^{\mathrm{i}x}+2\mathrm{i}\frac{\Omega'_{11}\Omega'_{12}}{\Omega_{21}'^2},\,x\in[-20.0,20.0],
\end{align}
where
\begin{align}\nonumber
\begin{split}
\Omega'_{11}=&\frac{1}{15625}\Big\{\big[(-1249\mathrm{i}-2498\mathrm{i}x-625.5-2502x^2-1251)\sqrt{1562501}-1559999\mathrm{i}-\\
&3129998\mathrm{i}x-781250.5-3125002x^2-1567501\big]\mathrm{e}^{\frac{-\mathrm{i}(-0.5-1250x)\sqrt{1562501}}{3125000}-\frac{\mathrm{i}x}{2}}\Big\}-\frac{4}{625}\bigg[\bigg(\\
&-0.5\mathrm{i}+\mathrm{i}x-0.25-x^2+\frac12\bigg)\sqrt{1562501}-\mathrm{i}+2500\mathrm{i}x+1251\bigg]\mathrm{e}^{\frac{\mathrm{i}(-0.5-1250x)\sqrt{1562501}}{3125000}-\frac{\mathrm{i}x}{2}},
\end{split}
\end{align}
\begin{small}
\begin{align}\nonumber
\begin{split}
\Omega'_{12}=&\frac{1}{781250}\Big\{\big[(-31250\mathrm{i}+62500\mathrm{i}-125000\mathrm{i}x^2-62500+125000x)\sqrt{1562501}-156250100\mathrm{i}\\
&(0.25+x^2)+78249950\mathrm{i}-78250050+156249900x\big]\mathrm{e}^{\frac{-\mathrm{i}(-0.5-1250x)\sqrt{1562501}}{3125000}+\frac{\mathrm{i}x}{2}}\Big\}+\frac{1}{390625}\\
&\bigg\{1249\bigg[\bigg(0.25\mathrm{i}+\mathrm{i}x^2+\frac{\mathrm{i}}{2}+\frac{-625.5}{1249}-\frac{1251x}{1249}\bigg)\sqrt{1562501}-\frac{1562501\mathrm{i}}{1249}(0.25+x^2)+\\
&\frac{4687499\mathrm{i}}{2498}+\frac{-2345000.5}{1249}+\frac{1559999x}{1249}\bigg]\mathrm{e}^{\frac{\mathrm{i}(-0.5-1250x)\sqrt{1562501}}{3125000}+\frac{\mathrm{i}x}{2}}\bigg\},
\end{split}
\end{align}
\end{small}
and
\begin{small}
\begin{align}\nonumber
\begin{split}
\Omega'_{21}=&\frac{1}{15625}\Big\{\big[(1249\mathrm{i}-625.5-2502x^2+2498\mathrm{i}x-1251)\sqrt{1562501}+1559999\mathrm{i}+3129998\mathrm{i}x-\\
&3125002(0.25+x^2)-1567501\big]\mathrm{e}^{\frac{\mathrm{i}(-0.5-1250x)\sqrt{1562501}}{3125000}+\frac{\mathrm{i}x}{2}}\Big\}+\frac{4}{625}\bigg[\bigg(-0.5\mathrm{i}+\mathrm{i}x+0.25+x^2\\
&-\frac12\bigg)\sqrt{1562501}-\mathrm{i}+2500\mathrm{i}x-1251\bigg]\mathrm{e}^{\frac{-\mathrm{i}(-0.5-1250x)\sqrt{1562501}}{3125000}+\frac{\mathrm{i}x}{2}}.
\end{split}
\end{align}
\end{small}

The Dirichlet boundary condition is
\begin{align}\label{E26}
q_{\mathrm{lb}}(t)=q_{\mathrm{rpw}}(-20.0,t),\,q_{\mathrm{ub}}(t)=q_{\mathrm{rpw}}(20.0,t),\,t\in[-0.5,0.5].
\end{align}
Here, applying the same data discretization method in Section 4.1, and  producing the training data which consists of initial data \eqref{E25} and boundary data \eqref{E26} by dividing the spatial region $[-20.0,20.0]$ into 513 points and the temporal region $[-0.5,0.5]$ into 401 points. We generate a smaller training dataset that containing initial-boundary data by randomly extracting $N_q=600$ from original dataset and $N_f=20000$ collocation points which are generated by the Latin hypercube sampling method. After giving a dataset of initial and boundary points, the latent rogue periodic wave solution $q(x,t)$ has been successfully learned by tuning all learnable parameters of the IPINN and regulating the loss function \eqref{E8}. The model of IPINN achieves a relative $\mathbb{L}_2$ error of 8.766380$\mathrm{e}-$02 in about $3862.2879$ seconds, and the number of iterations is 22414.

\begin{figure}[htbp]
\centering
\subfigure[]{
\begin{minipage}[t]{0.32\textwidth}
\centering
\includegraphics[height=3.5cm,width=4.8cm]{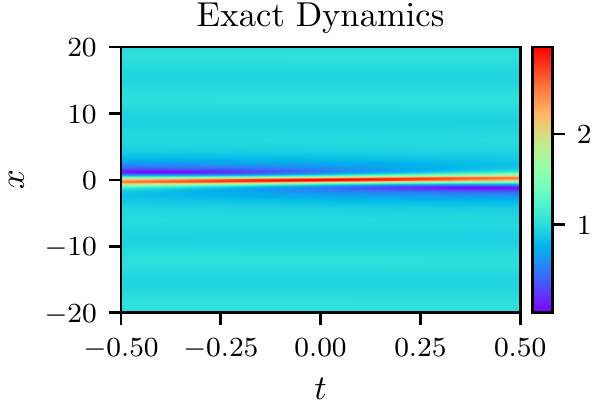}
\end{minipage}
}%
\subfigure[]{
\begin{minipage}[t]{0.32\textwidth}
\centering
\includegraphics[height=3.5cm,width=4.8cm]{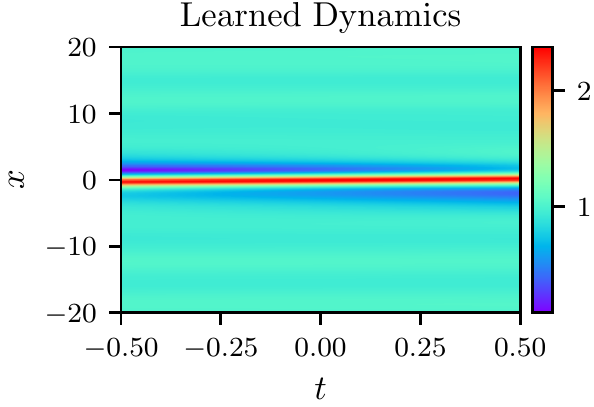}
\end{minipage}%
}%
\subfigure[]{
\begin{minipage}[t]{0.32\textwidth}
\centering
\includegraphics[height=3.5cm,width=4.8cm]{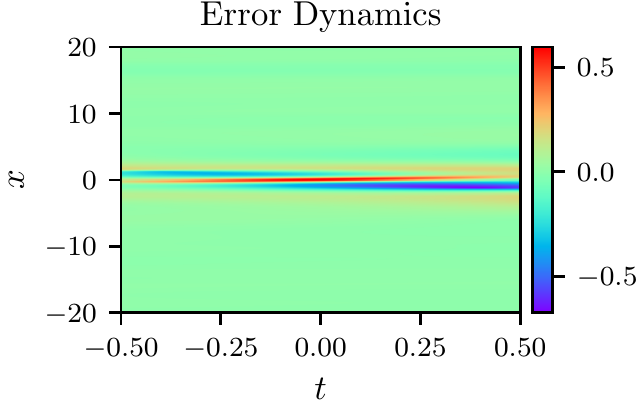}
\end{minipage}%
}%
\centering
\caption{(Color online) The rogue periodic wave solution $q(x,t)$ based on the IPINN: (a) The density plot of exact rogue periodic wave solution; (b) The density plot of learned rogue periodic wave solution; (c) The error density plot of the difference between exact and learned rogue periodic wave solution.}
\label{F14}
\end{figure}

\begin{figure}[htbp]
\centering
\subfigure[]{
\begin{minipage}[t]{0.32\textwidth}
\centering
\includegraphics[height=3.5cm,width=4.8cm]{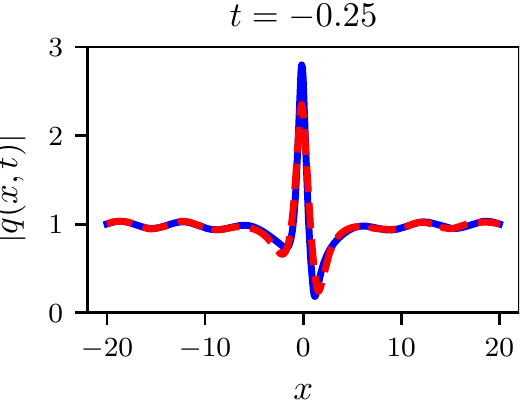}
\end{minipage}
}%
\subfigure[]{
\begin{minipage}[t]{0.32\textwidth}
\centering
\includegraphics[height=3.5cm,width=4.8cm]{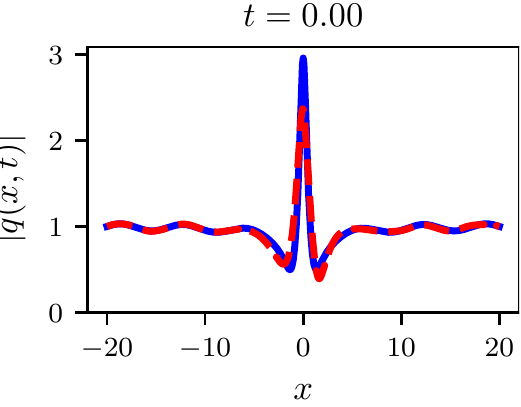}
\end{minipage}%
}%
\subfigure[]{
\begin{minipage}[t]{0.32\textwidth}
\centering
\includegraphics[height=3.5cm,width=4.8cm]{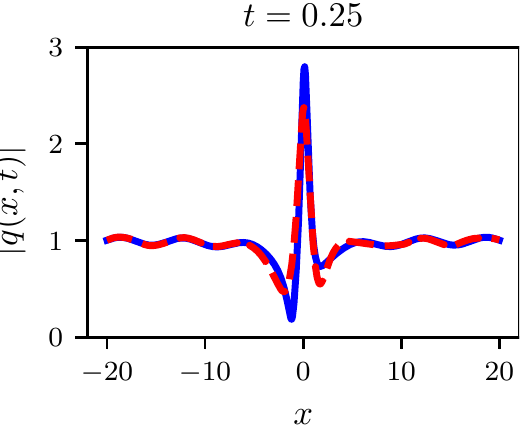}
\end{minipage}%
}%
\centering
\caption{(Color online) The sectional drawings of rogue periodic wave solution $q(x,t)$ based on the IPINN at (a): $t=-0.5$, (b): $t=0$ and (c): $t=0.5$.}
\label{F15}
\end{figure}

\begin{figure}[htbp]
\centering
\subfigure[]{
\begin{minipage}[t]{0.45\textwidth}
\centering
\includegraphics[height=5cm,width=6cm]{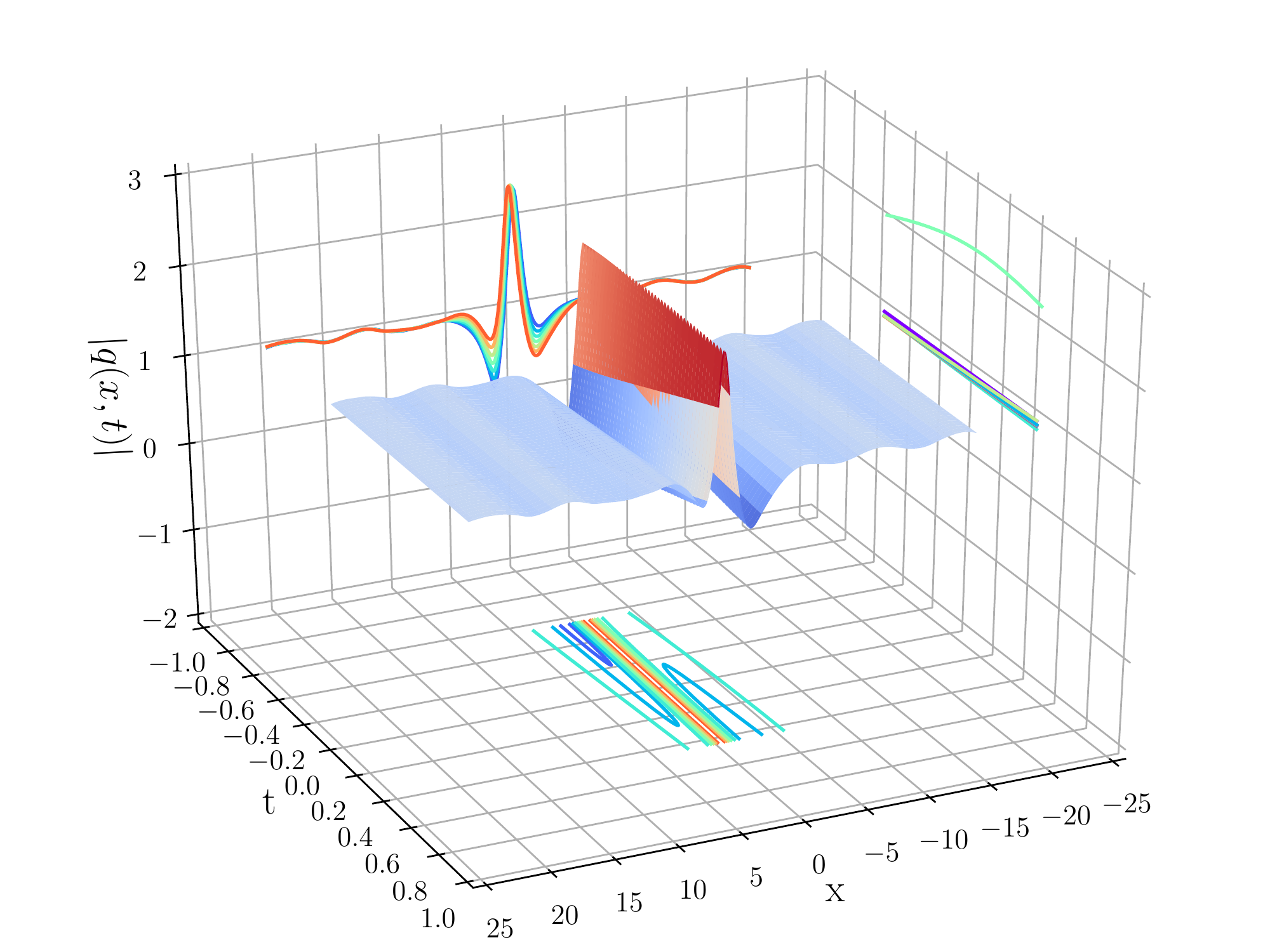}
\end{minipage}
}%
\subfigure[]{
\begin{minipage}[t]{0.45\textwidth}
\centering
\includegraphics[height=5cm,width=6cm]{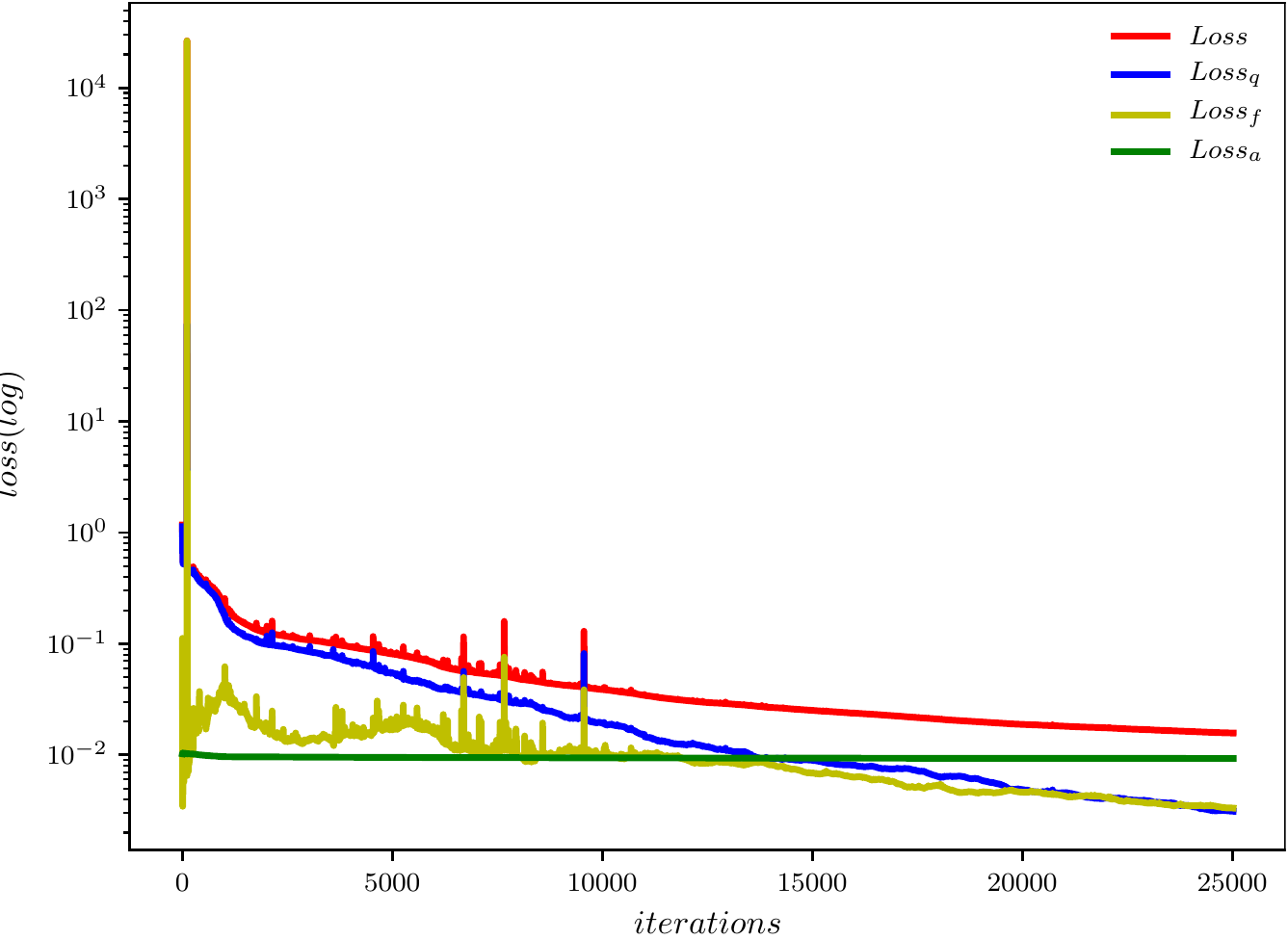}
\end{minipage}
}%
\centering
\caption{(Color online) The rogue periodic wave solution $q(x,t)$ based on the IPINN: (a) The three-dimensional plot; (b) The loss curve figure.}
\label{F16}
\end{figure}

In Figs. \ref{F14} - \ref{F16}, the density plots, the sectional drawing at different times and the iteration number curve plots for the rogue periodic wave solution $q(x,t)$ under IPINN structure are plotted respectively. Specifically, the density plots of exact dynamics, learned dynamics and error dynamics have exhibited in detail, and the corresponding peak scale is shown on the right side of the density plots in Fig. \ref{F14}. Specially, from the Fig. \ref{F14} (c), one can obviously find that the error range is about $-0.2$ to $0.2$. In Fig. \ref{F15}, we provide a comparison between the exact rogue periodic wave solution and the predicted solution based on the IPINN at different instants (a): $t=-0.25$, (b): $t=0$ and (c): $t=0.25$, and infer that the amplitude of the rogue wave solution reaches the maximum at time $t=0$, and the waveform of the profiles at time $t=-0.25$ and $t=0.25$ are symmetrical. The three-dimensional plot and its corresponding contour map of rogue periodic wave solution for the DNLS \eqref{E1} has been given out in the left panel (a) of Fig. \ref{F16}. From the right panel (b) of Fig. \ref{F16}, it is obvious that the $Loss$ curve (red solid line), $Loss_q$ curve (blue solid line) and $Loss_f$ curve (yellow solid line) arise oscillation before about 10000 iterations, especially $Loss_f$ curve oscillates most violently. After 10000 iterations, the three kinds of curves converge smoothly, and $Loss_q$ converges the fastest. Due to characteristics of $Loss_a$, the $Loss_a$ curve (green solid line) decreases steadily and slowly around $0.01$.

\section{Conclusion}
Increasing the performance of deep learning algorithms is significant in order to design fast and accurate machine learning techniques. An IPINN framework for extracting localized wave solutions dynamics of (1+1)-dimensional nonlinear time-dependent systems has been introduced from the spatiotemporal data. Specifically, we outline the flow-process diagram of DNLS equation based on the IPINN in detail. In this paper, we are committed to research the data-driven localized wave solutions which contain rational solution, soliton solution, periodic wave solution, rogue wave solution and rogue periodic wave solution for the DNLS by employing IPINN approach under the condition of small sample data set. The results show that the IPINN model could recover the different dynamical behaviors of localized wave solutions for DNLS fairly well.

Compared with the nonlinear Schr\"odinger equation and Chen-Lee-Liu equation, the DNLS has more nonlinear terms, so it is more difficult to recover data-driven solutions of the DNLS than the nonlinear Schr\"odinger equation and Chen-Lee-Liu equation \cite{Pu2021,PuJ2021,Peng2021}. As we can see from the Sections 3 and 4, it can be found that the $\mathbb{L}_2$ norm error can not reach the order of magnitude $\mathrm{e}-$03 by using IPINN method to recover the data-driven localized wave solutions with corresponding initial and boundary conditions. Moreover, we find that the data-driven localized wave solutions can not be recovered well when the temporal region $t$ is too wide. Therefore, we can find that more nonlinear terms have a greater impact on the performance of the neural network. Although higher order dispersion terms also have an impact on the performance of the neural network, the impact is not as violent as the nonlinear terms.

Compared with the classical PINN method, the influence of $Loss_q$ and $Loss_f$ on $Loss$ of IPINN approach is negligible due to the introduction of $Loss_a$. Generally speaking, from Sections 3 and 4, it is obvious that the curve of $loss_f$ fluctuates violently when the number of iterations is small, while the $Loss_a$ in all models decreases steadily around 0.01, which is determined by its slope mathematical structure. In the training process by employing IPINN method, the values of $Loss_q$ and $Loss_f$ are usually far less than the value of $Loss_a$, so their mathematical sum is mainly dominated according to $Loss_a$ and the Eq. \eqref{E8}, and $Loss_a$ is very stable , which ensures the overall topological stability of loss function. Although the values of $Loss_q$ and $Loss_f$ are larger than that of $Loss_a$ when simulating some complex solutions, for example, one can see that although the loss curve has obvious fluctuation in the early stage of training, the overall decline trend is stable and the decline is more and more stable after the iteration times are larger in Section 4.3. Apparently, $Loss_a$ plays an important role in the IPINN model, we can control the slope stability interval of $Loss$ by controlling the size of $N_a$. If $N_a$ is too small, the overall $Loss$ target value is too large. Conversely, if $N_a$ is too large, $Loss_a$ can not lead the loss function well, which will affect the stability of $Loss$ curve.

Compared with the traditional numerical methods, the IPINN has no limitation of grid size and large data set, and gives full play to the advantages of computer science and neural network. Moreover, the IPINN approach is trained with just few data, fast convergence speed and has a better physical interpretability by applying the physical constraints and locally adaptive activation function. The IPINN method showcases a series of results of various problems in the interdisciplinary field of applied mathematics and computational science, and opens a new path for using deep learning to simulate unknown solutions and correspondingly discover the parametric equations in scientific computing. It also provides a theoretical and practical basis for solving some high-dimensional scientific and big data space-time problems that can not be solved before. However, for the nonlinear integrable systems which contain high order nonlinear term is an unavoidable research hotspot. How to use the theory of integrable systems to improve IPINN model is a problem that needs further research in the future.

\section*{Acknowledgements}
\hspace{0.3cm}
The authors gratefully acknowledge the support of the Global Change Research Program of China (No.2015CB953904), the National Natural Science Foundation of China (No. 11675054), and Science and Technology Commission of Shanghai Municipality (No. 18dz2271000).

\end{document}